\documentclass[format=acmsmall, review=false, screen=true]{acmart}
\AtBeginDocument{%
  \providecommand\BibTeX{{%
    \normalfont B\kern-0.5em{\scshape i\kern-0.25em b}\kern-0.8em\TeX}}}

\setcopyright{acmcopyright}
\copyrightyear{2026}
\acmYear{2026}
\acmDOI{10.1145/xxxxxxx.xxxxxxx}

\acmJournal{TOSEM}
\acmVolume{0}
\acmNumber{0}
\acmArticle{0}
\acmMonth{0}

\acmISBN{978-1-4503-XXXX-X/18/06}

\usepackage{multirow}
\usepackage{colortbl}
\usepackage{rotating}
\usepackage{framed}
\usepackage{listings}
\usepackage{xcolor}
\usepackage{graphicx}

\usepackage{etoolbox}
\newif\ifcitationaudit
\citationauditfalse
\newcommand{\citefix}[1]{%
  \ifcitationaudit
    {\color{blue}\hypersetup{citecolor=blue,linkcolor=blue,urlcolor=blue}#1}%
  \else #1\fi}
\newcommand{\citationauditkeys}{}
\forcsvlist{\listadd\citationauditkeys}{corec,changescribe,unixcoder,codebert,codesumm2,zhang2024using,CoRaCMG,Zeng2026,Huang2026,Xiong2025C3GEN,Lin2023CCT5,He2023COME,li2024only,llmbeyond,bajracharya2010sourcerer,anquetil1999experiments,openai2022chatgpt,meta2024llama3,google2024gemma,vedantam2015cider,reimers2019sbert,wohlin2012experimentation,li2026ccmg,treesitter,jinacode,deepseekv32,deepseekv4,gpt54,beyer1999nearest,lan2026hisrag,mcinnes2018umap,wu2025icl,nngen,whatmakes,deltadoc,cmtgen,nmt,fira,autosumcc,cc2vec,codisum,ptrgncmsg,codet5,race,codet5p,plbart,bleu,rouge,meteor}
\makeatletter
\AtBeginDocument{%
  \let\citationaudit@lbibitem\@lbibitem
  \def\@lbibitem[#1]#2{%
    \normalcolor
    \citationaudit@lbibitem[#1]{#2}%
    \ifcitationaudit
      \ifinlist{#2}{\citationauditkeys}
        {\color{blue}\hypersetup{urlcolor=blue}}
        {\normalcolor}%
    \fi}}
\makeatother

\newif\ifcleanversion
\cleanversiontrue
\ifcleanversion
\newcommand{\rev}[1]{#1}

\newcommand{\todo}[1]{}

\else
\newcommand{\rev}[1]{\textcolor{blue}{#1}}

\newcommand{\todo}[1]{{\color{red}\textbf{[TODO: #1]}}}

\fi

\begin{document}

\title{CAMGen: A Core Change-Aware Multi-View Framework for Commit Message Generation}


\author{Bo Xiong}
\orcid{0009-0007-4559-7815}
\affiliation{%
  \institution{School of Computer Science, Wuhan University}
  \city{Wuhan}
  \country{China}}
\affiliation{%
  \institution{School of Computer Science, Peking University}
  \city{Beijing}
  \country{China}}
\email{yueshaomoon_@whu.edu.cn}

\author{Weisheng Chen}
\orcid{0009-0008-9817-5381}
\affiliation{%
  \institution{School of Computer Science, Wuhan University}
  \city{Wuhan}
  \country{China}}
\email{wisychan@whu.edu.cn}

\author{Zongen Ren}
\orcid{0009-0002-9924-543X}
\affiliation{%
  \institution{School of Computer Science, Wuhan University}
  \city{Wuhan}
  \country{China}}
\email{zongnren@whu.edu.cn}

\author{Yifeng Zhu}
\orcid{0009-0009-0487-8404}
\affiliation{%
  \institution{School of Computer Science, Peking University}
  \city{Beijing}
  \country{China}}
\email{yifengzhu25@stu.pku.edu.cn}

\author{Xutian Li}
\orcid{0009-0001-2441-7447}
\affiliation{%
  \institution{School of Computer Science, Peking University}
  \city{Beijing}
  \country{China}}
\email{xtli25@stu.pku.edu.cn}

\author{Runbang Yan}
\orcid{0009-0001-1465-6102}
\affiliation{%
  \institution{School of Computer Science, Peking University}
  \city{Beijing}
  \country{China}}
\email{rbyan25@stu.pku.edu.cn}

\author{Chong Wang}
\orcid{0000-0003-4576-5392}
\affiliation{%
  \institution{School of Computer Science, Wuhan University}
  \city{Wuhan}
  \country{China}}
\email{cwang@whu.edu.cn}

\author{Peng Liang}
\orcid{0000-0002-2056-5346}
\affiliation{%
  \institution{School of Computer Science, Wuhan University}
  \city{Wuhan}
  \country{China}}
\email{liangp@whu.edu.cn}
\renewcommand{\shortauthors}{Xiong et al.}


\begin{abstract}

  Commit messages help developers understand code changes and support software maintenance. Recent approaches use repository context to improve commit message generation (CMG) with large language models (LLMs). However, existing CMG methods provide limited support for understanding how local edits affect functionality jointly implemented by multiple code entities. This functional understanding is essential for generating commit messages that explain the main change and its impact. To address this problem, we propose \textbf{CAMGen}, a core \textbf{c}hange-\textbf{a}ware \textbf{m}ulti-view commit message \textbf{gen}eration framework that incorporates functional context into generation. CAMGen identifies core entities that represent the primary change intent. It then forms candidate functional units through repository-wide semantic clustering and uses those containing the core entities to construct a Functional Semantic View. This view supplies semantic context for interpreting the affected functionality. CAMGen also uses context provided by the Structural Dependency View and the Historical Exemplar View. The context from all three views is then provided to the LLM together with the code diff to guide CMG. 
  \rev{Experimental results across three programming languages show that CAMGen outperforms representative CMG baselines, including LLM-based approaches, on most automatic evaluation metrics.} Human evaluation further shows that CAMGen generates more complete commit messages with clearer explanations of change rationale and impact. These results demonstrate the value of functional context and complementary repository information for CMG. The replication package is available at~\cite{dataset}. 
\end{abstract}  

\ccsdesc[500]{Software and its engineering~Software maintenance tools}

\keywords{Commit Message Generation, Large Language Models, Semantic Clustering}



\maketitle

\section{Introduction}
Commit messages are essential software artifacts through which developers record, understand, and communicate code changes. Prior work commonly characterizes the core content of a high-quality commit message in terms of \textit{what} and \textit{why}: what summarizes the primary code modifications, whereas why explains the motivation or intended objective of the change~\citefix{\cite{whatmakes, Zeng2026}}. Together, these two types of information allow developers to understand a commit without inspecting the entire diff or tracing its complete history, thereby supporting code review, program comprehension, collaborative development, and long-term maintenance. Commit Message Generation (CMG) aims to automatically generate natural-language commit messages from code changes. Early approaches primarily rely on predefined templates\citefix{~\cite{deltadoc,changescribe}} or formulate CMG as a diff-to-text translation task\citefix{~\cite{cmtgen,nmt}}. Subsequent research uses the code understanding and natural-language generation capabilities of pretrained models and large language models (LLMs) while incorporating repository context, such as abstract syntax trees (ASTs), call graphs, dependency graphs, repository-level code fragments, and historical diff-message pairs\citefix{~\cite{fira,li2024only, Xiong2025C3GEN, CoRaCMG}}. These approaches supplement code diffs with structural, semantic, and historical information and have substantially improved the quality of generated commit messages.

Despite this progress, representative CMG methods still select repository context primarily according to its direct relevance to a diff. Structural approaches expand context along dependencies near changed locations, semantic retrieval approaches return code fragments with representations similar to the diff, and history-based approaches retrieve similar diff-message pairs. These approaches treat individual code fragments, structural relations, or historical exemplars as the basic units of context construction for the code diff query, but they do not systematically identify and organize the program entities that jointly implement the affected functionality. Since the implementation of software functionality often spans multiple entities distributed across different files or modules, context captured through local structural expansion or fragment-level similarity retrieval typically covers only a subset of the entities that implement the functionality affected by the diff. Such partial context cannot fully reveal the functional units affected by a diff or the responsibilities of these units within the system. 
The key \textbf{challenge} is therefore to help LLMs understand code diffs from a functional perspective so that they can generate commit messages that accurately describe the primary change and its functional impact. 

\begin{figure*}[h]
    \centering
    \includegraphics[width=\linewidth]
    {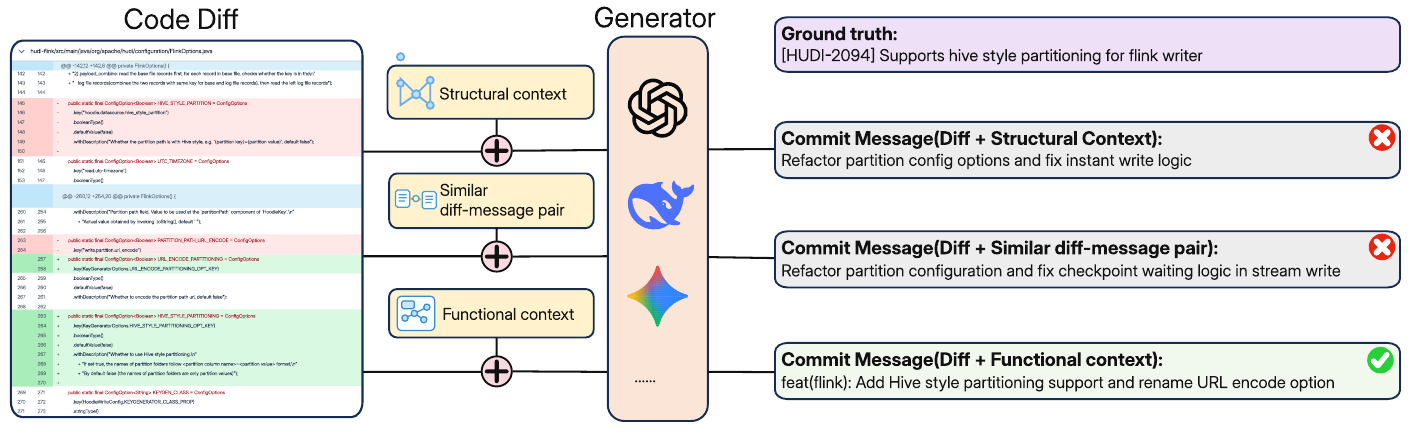}
    \par\smallskip
    \begin{minipage}{\linewidth}
        \footnotesize
        \raggedright
        \textsuperscript{*} Only part of the code diff is shown due to space constraints.
    \end{minipage}
    \caption{Motivating example of CMG with different context types.}
    \label{fig:motivation}
\end{figure*}

\textbf{Motivation.} Consider the scenario illustrated in Figure~\ref{fig:motivation}, which is drawn from commit \texttt{b8a8f57}\citefix{\footnote{\citefix{\url{https://github.com/apache/hudi/commit/b8a8f57}}}} in Apache Hudi project. The primary purpose of this commit is to add Hive-style partitioning support to the Flink writer, as expressed by the reference commit message written by the developer ``\textit{[HUDI-2094] Supports hive style partitioning for flink writer}''. The commit modifies multiple files. It introduces and propagates the Hive-style partitioning configuration, adjusts partition path parsing and reading logic, and extends the corresponding tests. Meanwhile, the diff also contains changes to configuration option names and local logic such as \texttt{instantToWrite}. These distributed and co-occurring edits make it difficult for the LLM to identify the primary functional intent of the commit. When the diff is supplemented with explicit structural dependency information, the LLM generates a commit message \texttt{Refactor partition config options and fix instant write logic}. This message identifies the changes to configuration options and write logic but still characterizes the commit as refactoring. Likewise, when the diff is supplemented with similar historical changes and their reference commit messages, the LLM generates \texttt{Refactor partition configuration and fix checkpoint waiting logic in stream write}. The historical context helps the model follow project-specific wording, but the generated message still focuses on local operations, such as configuration adjustments and checkpoint waiting logic, and fails to reveal the primary functional impact of adding Hive-style partitioning support. In contrast, when provided with context constructed through functional semantic clustering, the LLM generates \texttt{feat(flink): Add Hive style partitioning support and rename URL encode option}, which more closely reflects the primary intent expressed in the human-written commit message. This comparison between different context strategies shows that functionally cohesive repository context enables the model to associate relevant entities distributed across configuration declarations, parameter propagation, partition path processing, and test code with the same functional unit. The model can consequently understand the shared functional objective and behavioral impact of these local edits, thereby generating a commit message that accurately summarizes the primary modification and clearly explains its functional impact.

The limited ability of existing CMG methods to provide repository context that captures the functionality affected by code changes reduces the practical value of generated commit messages for developers and downstream software engineering tasks. Although a code diff records every edit in a commit, understanding the actual impact of these edits requires knowledge of the functional responsibilities of the affected entities and how the edits alter the execution logic or externally observable behavior of the corresponding functionality. Without functional context, an LLM may identify concrete operations, such as adding a condition check or adjusting a method call, but it may still fail to determine which functional unit is affected or what program behavior is changed. The resulting commit messages, therefore, tend to remain at the level of local operations or generic summaries and fail to provide complete and accurate descriptions of change impacts. Such quality limitations of commit messages force developers to inspect the code to determine the functional scope of a change, weakening the support provided by commit messages for code review, program comprehension, and software maintenance. Automated CMG, therefore, requires change-impact-aware repository context centered on functional context. Such context should connect the entities carrying the primary modification with related repository code, enabling the LLM to examine their functional responsibilities and the effects of the edits. This understanding supports commit messages that explain both the main modification and its functional and behavioral implications.

To address this challenge, we propose CAMGen, a core change-aware multi-view framework for commit message generation. Code modifications within a commit differ in their relevance to its primary purpose. Huang et al.~\cite{Huang2026} show that prioritizing core-class changes under a fixed input budget improves several CMG methods. This finding motivates focusing functional analysis on the main modification. CAMGen first identifies its core entities: classes or functions that carry the main logic modification, feature enhancement, or defect fix in the diff. Examining these entities together with functionally related entities helps reveal which functionality is affected and how the edits change its behavior. Core entity identification thus provides the starting point for understanding the diff from a functional perspective. Based on these core entities, CAMGen constructs complementary views corresponding to three types of information considered by developers when interpreting key changes: explicit structural relations between program entities, the functional roles of the changed entities within the system, and historical examples of similar changes. Specifically, CAMGen uses the core entities to construct a Structural Dependency View ($\mathrm{View}_{\mathrm{Struct}}$) and a Functional Semantic View ($\mathrm{View}_{\mathrm{Func}}$), while retrieving similar changes and their human-written commit messages from project history to construct a Historical Exemplar View ($\mathrm{View}_{\mathrm{Exemplar}}$). The $\mathrm{View}_{\mathrm{Struct}}$ captures explicit dependencies between the core entities and other program entities. The $\mathrm{View}_{\mathrm{Func}}$ organizes repository code into candidate functional units through semantic clustering. Following~\cite{Zhu2025RepoSummary}, we adopt method clusters as candidate functional features and select those containing the core entities to provide context for interpreting the affected functionality. The $\mathrm{View}_{\mathrm{Exemplar}}$ provides project-specific terminology and writing conventions. The $\mathrm{View}_{\mathrm{Func}}$ constitutes the central contribution of CAMGen: it helps the LLM relate the primary edits to the responsibilities of the changed entities and explain the functional impact of the code diff. By integrating local structural, global functional, and historical information, CAMGen enables LLMs to understand the effects of local code edits on system functionality and to generate accurate, comprehensive, and project-consistent commit messages.

The main \textbf{contributions} of this paper are as follows:
\begin{itemize}


    \item \textbf{Approach:} We propose CAMGen, a core change-aware multi-view CMG framework that constructs functional context through semantic clustering. It combines this context with structural dependencies and historical examples to guide LLMs in generating commit messages.
    \item \textbf{Evaluation:} 
    \rev{We evaluate CAMGen across three programming languages through automatic evaluation against seven representative CMG baselines, including LLM-based approaches, human evaluation, and ablation studies.}
     \item \textbf{Implication:} We provide implications for the effective integration and use of functional context in LLM-based CMG. This context helps LLMs understand the functional intent and impact of code changes and generate higher-quality commit messages.
\end{itemize}

\textbf{Organization}: Section~\ref{sec:related} reviews the related work. Section~\ref{sec:method} presents the proposed CAMGen framework. Section~\ref{sec:exp} describes the research design. Section~\ref{sec:5} presents and analyzes the experimental results. Section~\ref{sec:6} discusses threats to validity. Section~\ref{sec:conclusion} concludes the paper and outlines future work.

\section{Related Work and Background}
\label{sec:2}\label{sec:related}
In recent years, CMG has attracted considerable research attention. Existing studies have progressed from early rule-based and retrieval-based approaches to neural learning-based methods, and more recently to LLM-based methods that leverage the code-understanding and natural-language-generation capabilities of LLMs. We also discuss code clustering as a methodological background for understanding the functionality jointly implemented by related code entities.

\subsection{Commit Message Generation}
Early CMG approaches relied on predefined rules and templates~\cite{deltadoc,changescribe,autosumcc}. Subsequent research has advanced along four primary directions: retrieval-based, learning-based, hybrid, and LLM-based approaches~\cite{cmgreview}.

\textbf{Retrieval-based approaches.}
These approaches leverage commit messages associated with similar code changes. For example, NNGen~\cite{nngen} retrieves similar diffs using cosine similarity and BLEU, while CC2Vec~\cite{cc2vec} supports retrieval through learned code-change representations. The effectiveness of these approaches depends on the relevance of the retrieved code diffs, since the retrieved message is directly reused.

\textbf{Learning-based approaches.}
These approaches learn mappings from code diffs to commit messages. Early studies formulate CMG as a neural machine translation task, exploring attention and copying mechanisms to represent changes and handle code identifiers~\cite{cmtgen,nmt,codisum,ptrgncmsg}. Specifically, FIRA~\cite{fira} captures edit semantics through fine-grained graphs, while CodeT5~\cite{codet5} and CCT5~\cite{Lin2023CCT5} provide pre-trained representations for both code and code changes, respectively. IEEF~\cite{Chen2026IEEF} constructs high-quality training data and combines information extraction, information enhancement, and fine-tuning to improve the generation of \emph{what} and \emph{why} information. 
CCGen~\cite{Huang2026} estimates the importance of changed classes and prioritizes core-class changes in the generation input. Its controlled comparison on multi-class commits shows that this prioritization improves several CMG methods under a fixed token budget. These results suggest that CMG should pay more attention to the core modified entities in code diffs.

\textbf{Hybrid approaches.}
Hybrid approaches combine retrieval and generation. ATOM~\cite{ATOM} ranks retrieved and generated messages to select the final output, whereas CoRec~\cite{corec} and RACE~\cite{race} use similar historical commits to guide CMG. 
COME~\cite{He2023COME} uses modification embeddings to distinguish added, deleted, and unchanged tokens in a code diff and employs CodeT5 to generate a candidate commit message. It also retrieves a candidate commit message from a historical commit with a similar code diff, then uses a decision algorithm to select between the generated and retrieved commit messages. These methods incorporate historical examples into the commit message generation process.

\textbf{LLM-based approaches.}
Recently, some studies have investigated the use of LLMs for CMG through prompting and in-context learning~\cite{zhang2024using,lopes2024commitllm}. Wu et al.~\cite{wu2025icl} examined how in-context learning configurations affect generation quality and evaluated generalization across programming languages and newly collected commits.
To improve in-context example selection, Xue et al.~\cite{llmbeyond} proposed ERICommiter, which combines semantic and lexical retrieval to select in-context examples efficiently. CoRaCMG~\cite{CoRaCMG} uses hybrid retrieval of historical diff--message pairs to augment both fine-tuned models and LLM prompts. HisRag~\cite{lan2026hisrag} extends retrieval to developer and repository histories, providing historical messages as generation context.
Researchers have also explored supplementing the diff with broader software context to help LLMs interpret code changes into commit messages.  OMG~\cite{li2024only} obtains software context through tool invocation. OMEGA~\cite{Imani2025OMEGA} refines summaries of changed methods and explanations of the diff to support CMG with a compact local LLM. C3Gen~\cite{Xiong2025C3GEN} retrieves contextual code using the repository structure. CCMG~\cite{li2026ccmg} combines project-agnostic examples with project-specific information at the project, code, and style levels, and uses two-stage prompting to generate conventional commit messages. 
CMO~\cite{Li2026CMO} uses an LLM to generate revised candidate commit messages based on a human-written commit message, the code diff, and additional software context. LLM-based and retrieval-based quality evaluators assess these candidate commit messages, and the highest-scoring candidate is selected for further refinement.

\textbf{Empirical studies and evaluation.}
Empirical studies on CMG assess the performance of existing generation methods and investigate whether evaluation metrics accurately reflect the quality of generated commit messages. Wang et al.~\cite{Wang2025Holistic} evaluated seven CMG methods on the HORDA dataset, which extends CMG evaluation to commits with long code diffs or long commit messages. They found that a retrieval-based CMG method outperformed non-pretrained neural generators in within-project evaluation, while pretrained generators performed better than non-pretrained ones in CMG. Huang et al.~\cite{Huang2024ExplicitImplicit} evaluated learning-based CMG methods on two types of target commit messages: explicit messages that directly summarize code edits and implicit messages that describe the overall nature or intent of a change. They found that the methods performed worse when generating implicit commit messages and generalized poorly across the two commit message types. 
Tsvetkov et al.~\cite{Tsvetkov2025Realistic} investigated whether automatic CMG evaluation metrics reflect the amount of textual editing needed to make commit messages generated by GPT-4 suitable for submission to version control. They constructed an evaluation dataset by pairing the generated commit messages with the corresponding commit messages edited by experts. Using these expert editing examples, they prompted GPT-4 Turbo to generate additional pairs of original and edited commit messages. For each pair, they used edit distance to quantify the amount of textual editing. They then examined the correlation between this amount and automatic metric scores calculated against independent reference commit messages for the same commit. The Edit Distance metric shows the strongest correlation with the amount of textual editing, whereas BLEU and METEOR scores are only weakly correlated with it.
Kumar et al.~\cite{Kumar2025Evaluators} investigated whether LLMs could evaluate commit message quality as an alternative to text similarity metrics. Their experiments with GPT-4o, LLaMA 3.1, and Mistral Large show that LLM assessments of commit message quality align with human evaluations of the quality of the same commit messages. Zeng et al.~\cite{Zeng2026} investigated the use of LLMs to evaluate whether generated commit messages accurately describe code changes and explain the rationale behind those changes. Using GPT-4 with chain-of-thought and few-shot prompting, the study obtained assessments that agreed more closely with human ratings of both aspects of commit message quality than the textual and semantic similarity metrics examined.

\subsection{Code Clustering for Program Comprehension}
\label{subsec:related-clustering}
Code clustering organizes repository code into candidate functional units, which are groups of code entities that have related responsibilities or contribute to a common functionality across files~\cite{Zhu2025RepoSummary}. Such groupings support program comprehension by bringing related entities into a shared context for examining their roles in the system and the functionality they support~\cite{kuhn2007semantic}.

Kuhn et al.~\cite{kuhn2007semantic} use latent semantic indexing to measure similarities in identifiers and comments, then cluster source artifacts into linguistic topics. Labels and visualizations of these topics reveal where related concepts occur across the existing module structure. SARIF~\cite{Zhang2023SARIF} combines dependencies, topics extracted from code text, and folder structure in a weighted graph, then clusters source files to recover architectural components. These studies show how clustering exposes relationships among code elements beyond their physical organization in a repository.

More recent work uses clustering to generate documentation that explains repository functionality. HGEN~\cite{Dearstyne2024HGEN} clusters embeddings of code summaries using multiple clustering algorithms, then generates descriptions of features or user stories from the selected clusters. It iteratively builds higher levels of documentation and establishes traceability links between the generated descriptions and the underlying artifacts.

\rev{FeatOutline}~\cite{Zhu2025RepoSummary} further organizes code around functional features through clustering at both the file and method levels. It extracts import and invocation dependencies and generates summaries of files and methods. Similarities between summary embeddings are combined with the corresponding dependencies to form weighted graphs. Using these graphs, it applies Leiden community detection first to files and then to methods within each file group, with the aim of bringing together methods with similar responsibilities or contributions to a common feature. The resulting method clusters are treated as candidate functional features. An LLM uses the method summaries in each cluster to generate feature descriptions, which are then organized into higher-level documentation. Traceability links connect the generated features to their associated methods and files, allowing developers to navigate from a functional description to its implementation. \rev{Its evaluation reports more complete coverage of documented functionality and more accurate feature-to-code traceability than the repository summarization baselines, providing empirical support for feature-oriented code organization.}

These studies establish clustering as a means of organizing code for program comprehension and functional documentation. This suggests a way to support CMG: related repository code can provide context for interpreting how code diffs affect functionality and behavior. Such context helps the LLM connect concrete modifications to the purpose and impact that an informative commit message should convey.

\subsection{Conclusive Summary}
\label{subsec:conclusive-summary}
Existing studies have leveraged code clustering to organize repository entities into functional groups, thereby facilitating program comprehension and documentation generation. Meanwhile, current CMG approaches incorporate historical examples and repository context to assist models in interpreting code diffs. However, these methods primarily retrieve similar historical diff-message pairs or code fragments directly related to the current diff, without explicitly linking the code diff to the broader functionality jointly implemented by the changed entities and other functionally related code entities. Such information about this broader functionality is crucial for enabling LLMs to comprehend the impact of code diffs, ultimately generating commit messages that accurately describe both the code diffs and their functional implications.

Our proposed framework, CAMGen, addresses this gap by enabling LLMs to understand code diffs from a functional perspective. CAMGen first identifies the core entities carrying the main modifications. It then constructs candidate functional units through semantic clustering and extracts context from those containing the core entities. Structural dependencies around the same core entities clarify explicit code relationships, while historical diff--message examples provide references for project-specific terminologies and writing styles. CAMGen provides these contexts together with the code diff to guide the LLM in generating commit messages that capture the primary change and its functional impact.

\section{Methodology}
\label{sec:methodology}\label{sec:method}
This section presents CAMGen, a three-stage framework for generating commit messages with repository context. First, core entity identification locates the classes or functions carrying the main modification in the code diff and resolves them to repository entities. Second, CAMGen constructs the Structural Dependency View ($\mathrm{View}_{\mathrm{Struct}}$) and Functional Semantic View ($\mathrm{View}_{\mathrm{Func}}$) around these entities, and retrieves historical diff--message examples using the current code diff to form the Historical Exemplar View ($\mathrm{View}_{\mathrm{Exemplar}}$). Finally, an LLM generates the commit message from the raw diff and the combined context.

\subsection{Overview}
\begin{figure*}[t]
    \centering
    \includegraphics[width=\linewidth]{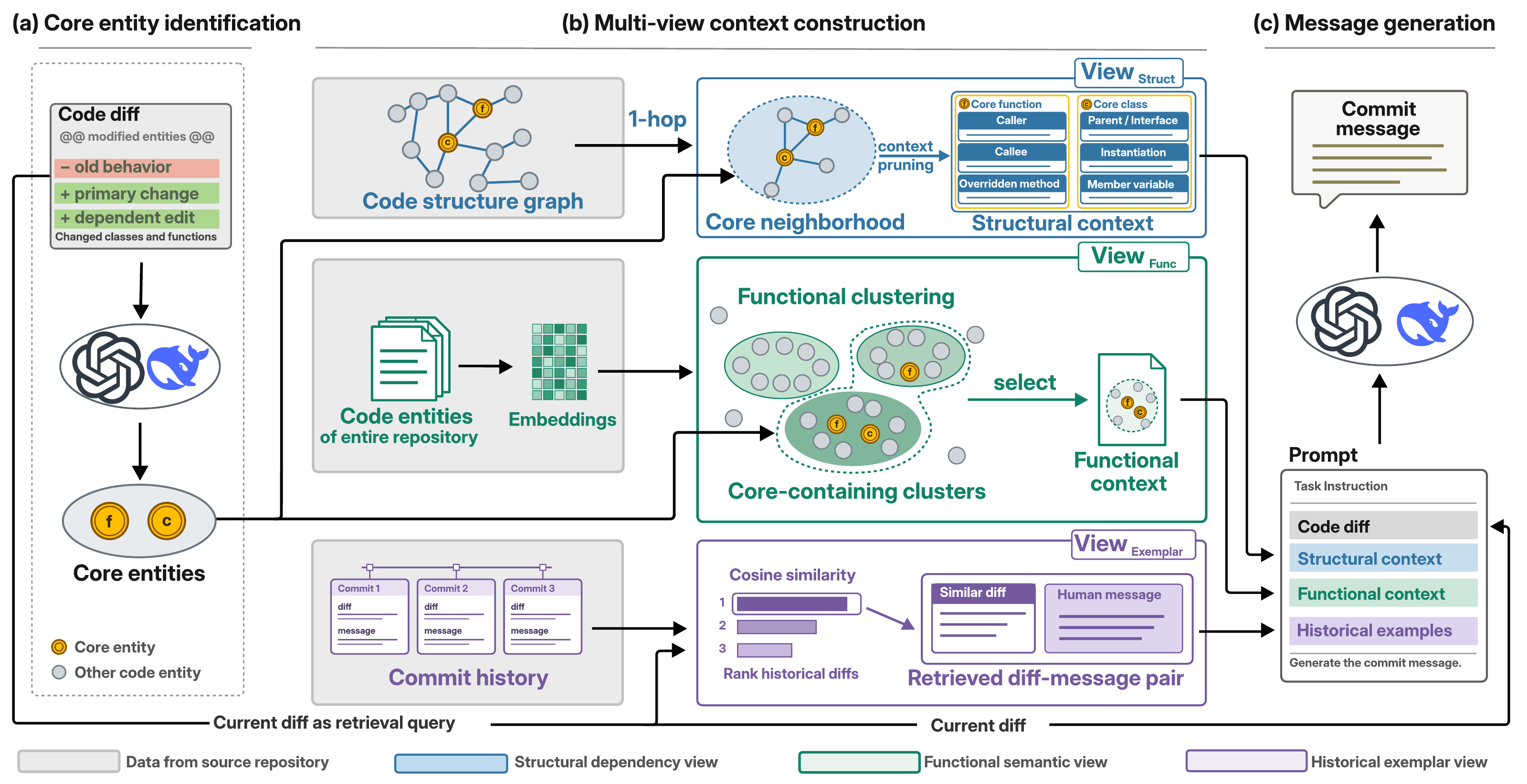}
    \caption{Overview of the \textsc{CAMGen} framework. 
    Given a code diff, \textsc{CAMGen} first performs core entity identification, and then constructs three complementary views, namely the $\mathrm{View}_{\mathrm{Struct}}$, the $\mathrm{View}_{\mathrm{Func}}$, and the $\mathrm{View}_{\mathrm{Exemplar}}$. 
    The three views are integrated to guide CMG.}
    \label{fig:framework}
\end{figure*}

Figure~\ref{fig:framework} presents an overview of \textsc{CAMGen}, a three-stage framework that leverages LLMs to comprehend the intent and functional impact of code changes for the automatic generation of high-quality commit messages.  
More specifically, the \emph{Core entity identification} stage locates the classes or functions carrying the main modification in the code diff. These entities focus the subsequent analysis on the affected functionality: their dependencies and functionally related entities help explain how the code diff changes program behavior. Then, three complementary views are constructed in the stage of \emph{Multi-view context construction} to capture the change context from different perspectives: the $\mathrm{View}_{\mathrm{Struct}}$ models explicit static dependencies derived from core entities, the $\mathrm{View}_{\mathrm{Func}}$ retrieves semantically related entities beyond one-hop dependencies, and the \emph{Historical Exemplar View} ($\mathrm{View}_{\mathrm{Exemplar}}$) incorporates project-specific terminologies and writing styles from historical commits. Finally, the context provided by the three views is combined with the code diff and fed into the LLMs in the \emph{Message generation} stage to produce commit messages that are structurally grounded, semantically informative, and stylistically consistent with project-specific conventions. 

\subsection{Core Entity Identification}\label{subsec:core-entity}
To precisely anchor the intent of a code change, we use an LLM as an intent identifier to analyze the raw diff and locate the entities carrying the primary logic changes, which are referred to as \emph{core entities} (i.e., core classes or core functions) in this paper. In this stage, we design a structured prompting strategy. Specifically, the system prompt constrains the LLM to generate outputs strictly in JSON format, while the user prompt explicitly defines \emph{core entities} as the primary carriers of key logic modifications, functional enhancements, or defect fixes in the code change.

Since the entity names identified by the LLM are typically in simplified form (e.g., class names or function names), whereas entities in the repository are identified by fully qualified names, an entity name resolution step is required. We employ a suffix-based robust matching algorithm to map simplified names to concrete entity nodes in the repository. The resolved entities serve as shared inputs to both the $\mathrm{View}_{\mathrm{Struct}}$ and the $\mathrm{View}_{\mathrm{Func}}$. This design decouples intent anchoring from context construction, avoiding redundant entity identification while reducing inconsistency and computational overhead.

\subsection{Structural Dependency View ($\mathrm{View}_{\mathrm{Struct}}$)}\label{subsec:structural-view}
\rev{The construction of $\mathrm{View}_{\mathrm{Struct}}$ builds upon the repository-context retrieval approach introduced in our previous work C3Gen~\cite{Xiong2025C3GEN}. We adapt its structural retrieval by using the core entities identified in Section~\ref{subsec:core-entity} as anchors.} This view employs static analysis to capture explicit structural dependencies within the code repository and extracts relevant contextual code with a high signal-to-noise ratio. The construction process consists of two steps: code-structure graph construction and context pruning.

\paragraph{Code Structure Graph Construction.}
First, we conduct static analysis over the entire code repository, taking the identified core entities as starting points, to construct a \emph{Code Structure Graph} (CSG). Nodes in the CSG represent code entities at multiple granularities, including directories, files, classes, interfaces, and functions. Edges encode three types of essential semantic relationships: (i) \emph{Call} edges, which capture invocation dependencies between functions; (ii) \emph{Implement/Extend} edges, which model inheritance or implementation relationships between classes and interfaces; and (iii) \emph{Instantiate} edges, which indicate locations where classes are instantiated. The resulting graph models the repository-level topology of explicit structural dependencies.

\paragraph{Context Pruning.}
To mitigate the inclusion of excessive irrelevant code, prior work has explored various strategies for selecting and pruning repository context. For example, COMMIT employs program slicing to retain statements related to code changes~\cite{jiang2024commit}. RepoHyper filters the graph expansion paths and reranks candidate nodes~\cite{phan2025repohyper}. HCP prunes cross-file context based on file dependencies and function-level relevance~\cite{zhang2025hcp}. In CAMGen, we design a role-aware pruning strategy. Specifically, we restrict graph traversal to one hop from each core entity and selectively retain neighboring nodes based on their semantic roles relative to the core entity.

When the core entity is a \emph{function}: (i) for downstream callees, we retain only the function skeleton, including the signature, parameters, return type, and documentation string, to convey functional contracts rather than implementation details; (ii) for upstream callers, we extract the minimal logical block enclosing the call site to highlight how the change affects external invocations; and (iii) if the function overrides a method in a parent class, we extract the corresponding method skeleton from the parent class to facilitate behavioral difference analysis.

When the core entity is a \emph{class}: (i) for its parent classes or implemented interfaces, we retain the class skeleton (i.e., the set of method signatures) to expose inheritance contracts; (ii) for instantiation sites, we preserve only the minimal context snippet containing the instantiation statement; and (iii) for newly added or modified member variables, we retrieve their type definitions and extract type summaries, covering the class name and public method signatures, to clarify the semantic origin of the newly introduced state.

This pruning strategy preserves essential structural dependencies while substantially reducing context size, thereby providing high-density and structured inputs for subsequent CMG.

\subsection{Functional Semantic View ($\mathrm{View}_{\mathrm{Func}}$)}
\label{subsec:functional-semantic-view}
The $\mathrm{View}_{\mathrm{Func}}$ provides repository context for interpreting the functional purpose and impact of a code diff. Following prior work~\cite{Zhu2025RepoSummary}, \rev{we use semantic proximity as a heuristic to form \emph{candidate functional units}} and treat clusters of semantically similar code entities as candidate functional units, which are groups of methods that may share related responsibilities or contribute to a common functionality. From these clusters, we select those containing the core entities identified in Section~\ref{subsec:core-entity} and construct the functional context by aggregating all code entities and their source code within the selected clusters. The resulting code context is supplied together with the diff to support functional interpretation during CMG. This process consists of three steps: semantic vectorization, semantic clustering, and cluster refinement.

\paragraph{Semantic Vectorization.}\label{sec:semantic-vectorization}
Semantic vectorization provides the representations used to group repository entities into candidate functional units. We use CodeBERT~\citefix{\cite{codebert}}, a pre-trained code language model, to encode the entities. Before encoding, we prepare the input code to account for the 512-token input limit of CodeBERT. For \emph{class} entities exceeding this threshold, we apply Tree-sitter-based \emph{code skeleton extraction}~\cite{treesitter}, preserving the class declaration, method signatures, and complete field definitions while removing method implementation bodies. For class entities within the token budget and all non-class entities (e.g., functions), we retain the original code. This preparation preserves structural and semantic cues from large classes while reducing the amount of code subject to truncation. We encode each prepared snippet into a 768-dimensional vector and normalize it to unit length. \citefix{In high-dimensional spaces, pairwise distances tend to be less distinguishable, reducing the contrast between near and far neighbors~\cite{beyer1999nearest}.} We therefore project the normalized vectors into a 20-dimensional space using UMAP~\cite{mcinnes2018umap}, which seeks to preserve local neighborhood structure. These projected representations serve as input to the clustering step.

\paragraph{Semantic Clustering.}\label{sec:semantic-clustering}
We apply HDBSCAN~\cite{campello2015hdbscan} to the projected representations to form candidate functional units. The resulting clusters range from a few to dozens of entities, while some entities may be labeled as noise. To preserve all core entities during context construction, we reassign any core entity labeled as noise to the nearest non-noise cluster. We then retain clusters containing at least one mapped core entity, focusing the selected context on the primary change. Formally, let $\mathcal{C} = \{C_1, C_2, \ldots, C_m\}$ denote the clusters after this reassignment and $\mathcal{E}_{mapped}$ denote the set of mapped core entities. The selected candidate functional units are defined as: $$\mathcal{C}_{core} = \left\{C_i \in \mathcal{C} \mid C_i \cap \mathcal{E}_{mapped} \neq \emptyset\right\}$$

\paragraph{Cluster Refinement.}\label{sec:cluster-refinement}
The goal of this step is to keep each selected cluster compact enough for CMG while preserving the core entities and \rev{nearby entities in the semantic embedding space}. Although HDBSCAN produces semantically cohesive groups, some clusters may still contain too many entities to serve as concise context. To this end, we refine each cluster in $\mathcal{C}_{core}$ by imposing a core-centric cluster-size limit $M$. Specifically, for any cluster containing more than $M$ entities, we retain all core entities and fill the remaining slots with non-core entities closest to the core centroid in semantic space. Clusters containing at most $M$ entities are kept unchanged. We set $M{=}10$ by default as a practical trade-off between contextual coverage and prompt budget. Each retained entity contributes a non-trivial amount of source code, and the prompt must also reserve space for the code diff and task instructions. A larger $M$ increases token cost and semantic noise, whereas a smaller $M$ may omit useful functional neighbors of the core entities. \rev{This refinement produces compact candidate context centered on the core entities for subsequent CMG.}

\subsection{Historical Exemplar View ($\mathrm{View}_{\mathrm{Exemplar}}$)}\label{subsec:historical-view}
While $\mathrm{View}_{\mathrm{Struct}}$ and $\mathrm{View}_{\mathrm{Func}}$ characterize the structural and semantic context of code changes from a static perspective, they are insufficient to capture project-specific terminologies and writing styles of commit messages. In practice, semantically similar changes may be described differently across projects. For example, fixing a null pointer exception may be phrased as ``fix NPE in X'' in one project, whereas another may prefer ``prevent null dereference in Y''. 
\rev{Following the historical-exemplar augmentation principle of our previous work CoRaCMG~\cite{CoRaCMG}, we incorporate the \emph{Historical Exemplar View} ($\mathrm{View}_{\mathrm{Exemplar}}$), which provides project-specific terminology and writing styles by retrieving the similar diff-message pair from the project's commit history.}

This view is constructed via \rev{hybrid lexical and semantic retrieval} to efficiently identify relevant historical exemplars. Specifically, we treat the diff of the current commit as a query and encode it into a 768-dimensional dense vector using a code-specific embedding model trained by Jina AI (\texttt{jina-embeddings-v2-base-code})\citefix{~\cite{jinacode}}. Pre-trained on large-scale, multi-language code corpora, the model can capture the semantic essence of code changes (such as functional intent and modification patterns), rather than relying solely on surface-level textual or structural features. \rev{In parallel, we apply BM25~\cite{bm25} to compute lexical relevance scores between the query code diff and candidate historical diffs.}


To enable efficient retrieval, we precompute embeddings for the diffs in the retrieval corpus and apply L2 normalization, so that similarity computation can be reduced to efficient dot-product operations. For each target commit, we restrict the retrieval candidates to commits from the same repository whose timestamps are strictly earlier than that of \rev{the commit corresponding to the query code diff}. \rev{We then compute cosine similarity between the query vector and the vectors of these eligible commits. Following CoRaCMG~\cite{CoRaCMG}, we normalize the BM25 scores and cosine similarity scores of these eligible commits to a common scale and combine them with equal weights (1:1). We rank the candidates in descending order of the fused score, select the highest-ranked candidates, and extract their corresponding diffs and developer-written commit messages as \emph{historical diff-message pairs} (exemplars).} These exemplars reflect how developers describe semantically similar changes in practice, thereby providing direct references to project-specific terminologies and writing styles for CMG.

Finally, the selected exemplars are integrated as supplementary context and jointly fed into an LLM in the \emph{Message generation} stage with the outputs from the other two views. The key value of this view lies in its ability to convey not only historical code diff-message pairs but also project-specific terminologies and writing styles. These historical examples guide the LLM in describing the current change using project-specific terminology and writing conventions.

\subsection{Commit Message Generation}

In this stage, CAMGen combines the raw code diff with the context from \(\mathrm{View}_{\mathrm{Struct}}\), \(\mathrm{View}_{\mathrm{Func}}\), and \(\mathrm{View}_{\mathrm{Exemplar}}\) into a unified prompt for an LLM. The structural context captures structural dependencies around the core entities of the code diff. The functional context complements these dependencies with code from the selected candidate functional units, helping the LLM interpret the responsibilities of the changed entities and the behavioral impact of the diff. Historical diff–message pairs provide examples of project-specific terminologies and writing conventions. 
The LLM then uses the code diff and the multi-view context to generate the corresponding commit message.

\section{Research Design}\label{sec:4}\label{4-research-design}\label{sec:exp}

\subsection{Research Questions}
The main objective of this study is to systematically evaluate the effectiveness of CAMGen for CMG by answering four research questions (RQs). More specifically, 
RQ1 evaluates the overall performance of the CAMGen framework by comparing it with representative CMG baselines using automatic evaluation metrics. Based on these results, RQ2 investigates the specific components that contribute to its performance gains. 
RQ3 further assesses the quality of the generated commit messages through human evaluation from the developers' perspective. Finally, RQ4 conducts a case study of representative CAMGen outputs with low automatic metric scores to identify the causes of these low automatic evaluation scores.

\textbf{RQ1: Overall Effectiveness: How does CAMGen perform compared to representative SOTA CMG approaches? }

This RQ aims to investigate the overall effectiveness of our CAMGen framework by evaluating the quality of generated commit messages with automatic metrics.
\rev{We compare CAMGen with six existing CMG approaches, i.e., NNGen, CoRec, RACE, COME, CCT5, and OMEGA, together with a diff-only LLM baseline, which directly prompts an LLM to generate a commit message from the code diff, without additional repository context. OMEGA and the diff-only baseline both use DeepSeek-V3.2, enabling comparisons with CAMGen under the same generator.}
To examine the influence of the underlying LLM, we employ DeepSeek-V3.2, DeepSeek-V4-Flash, and GPT-5.4 as the backbone generators in independent experiments.

\textbf{RQ2: Ablation Study: How do the individual design components of CAMGen contribute to its performance in CMG?}
To systematically evaluate the contribution of each design component within the CAMGen framework, we decompose this RQ into two sub-RQs:

\begin{itemize}
\item
  \textbf{RQ2.1: Core Entity Identification: Can CAMGen reliably identify and anchor core entities to serve as the foundational input for change-aware context construction?}
\item
  \textbf{RQ2.2: View Contribution Analysis: What are the individual and combined contributions of $\mathrm{View}_{\mathrm{Struct}}$, $\mathrm{View}_{\mathrm{Func}}$, and $\mathrm{View}_{\mathrm{Exemplar}}$ to the overall performance of CAMGen?}
\end{itemize}

More specifically, RQ2.1 is designed to examine whether the core entities identified by an LLM provide reliable anchors for subsequent context construction. 
RQ2.2 progressively incorporates the contexts provided by $\mathrm{View}_{\mathrm{Struct}}$, $\mathrm{View}_{\mathrm{Func}}$, and $\mathrm{View}_{\mathrm{Exemplar}}$ on the same samples. Using a comparison of the Diff-only baseline and seven CAMGen configurations incorporating single-view, two-view, or three-view context, this ablation study intends to evaluate the individual contributions and combined complementarity of each view. 
The answers to these two sub-RQs would help pinpoint the sources of CAMGen's performance gains and provide empirical evidence validating the integration of the change-aware context from three perspectives.

\textbf{RQ3: Human Evaluation: How does CAMGen perform from the developers' perspective?}
RQ3 is designed to assess whether the improvements captured by automatic metrics correspond to the improvement in the perceived quality of the generated commit messages. Automatic metrics primarily measure the lexical, sequential, or semantic similarity between a generated message and a reference commit message, and may fail to fully reflect whether the message accurately captures the core changes, explains its underlying rationale, or describes its corresponding functional impact. Building on these limitations, RQ3 empirically investigates the practical quality of commit messages generated by CAMGen from the developers' perspective, with dedicated attention to \textit{Comprehensiveness}, \textit{Rationality}, \textit{Change Impact Explanation}, \textit{Conciseness}, and \textit{Expressiveness}. Answering this RQ will not only help evaluate the practical value of CAMGen for developers, but also verify whether the performance gains observed in automatic metrics align with improvements in perceived commit message quality.

\textbf{RQ4: Qualitative Case Analysis: What are the underlying causes of CAMGen's underperformance under automatic metrics?}
This RQ aims to conduct a qualitative analysis of CAMGen's underperforming cases. Specifically, we sample instances from RQ1 where generated commit messages of CAMGen receive low scores on automatic metrics, and analyze these cases to identify and summarize the categories of poor performance. Answering this RQ would shed light on the factors that limit the performance of CAMGen and offer guidance for further improvements to CMG.

\subsection{Research Datasets}\label{sec:research-datasets}
All experiments in this study are conducted on ApacheCM\citefix{~\cite{Xiong2025C3GEN}}, a dataset constructed from open-source projects maintained by the Apache Software Foundation. Each commit record contains the code diff, commit message, repository identifier, commit hash, author, commit date, modified files, and additional metadata. Unlike datasets that retain only \texttt{\textless{}diff,\ message\textgreater{}} pairs, ApacheCM preserves this richer metadata, enabling data partitioning based on programming language, repository-level analysis, and downstream context retrieval.

To address RQ1, we construct evaluation subsets for Java, C++, and Python to assess the across-language performance of CAMGen. For each programming language, we randomly sampled 3,000 commits from ApacheCM to construct an initial candidate set. This common sampling budget keeps the candidate sets comparable in size while allowing for exclusions during preprocessing. We then exclude commits whose corresponding repository snapshots cannot be recovered or for which the static analysis required for context construction cannot be completed. After this screening, the final evaluation set contains 2,498 C++ commits, 2,469 Java commits, and 2,555 Python commits, totaling 7,522 commits. 

To investigate the effectiveness of LLM-based core-entity identification in CAMGen, we randomly sample 100 commits from ApacheCM for RQ2.1. We manually annotate program entities, including classes, functions, and interfaces, and treat these annotations as the gold labels. DeepSeek-V4-Flash and DeepSeek-V4-Pro then identify core entities in the same commits. 

To analyze individual contributions, combined effects, and complementarity of the three views, we randomly sample 1,000 commits from the three language datasets used in RQ1 for the RQ2.2 ablation study. We employ DeepSeek-V3.2 as the generator and compare \rev{the Diff-only baseline with seven CAMGen configurations: three single-view configurations, three two-view configurations, and CAMGen (Full). The baseline and all seven CAMGen configurations take the raw code diff as input. The Diff only baseline directly prompts the LLM to generate commit messages based solely on the code diff.} The three single-view configurations add context provided by $\mathrm{View}_{\mathrm{Struct}}$, $\mathrm{View}_{\mathrm{Func}}$, and $\mathrm{View}_{\mathrm{Exemplar}}$, respectively. The two-view configurations combine context provided by two views, whereas CAMGen (Full) uses context from all three views.

For human evaluation in RQ3, we sample 50 commits from the RQ2.2 ablation subset and reuse the corresponding DeepSeek-V3.2 outputs for the eight context configurations. Each evaluation instance includes the code diff, the reference commit message, the messages generated by the baselines and CAMGen configurations, and access to browsable and searchable repository snapshots from before and after the commit. Three human evaluators with software development experience independently score the messages without knowing their sources. For each method or configuration, we first average the final individual ratings of the three evaluators within each commit and then average these scores across the 50 commits.

Table~\ref{tab:dataset-rq-overview} summarizes the evaluation sets used to answer RQ1-RQ3. The table reports the number of commits, the sampling frame, and the corresponding research purpose. 

\begin{table}[t]
    \centering
    \caption{Overview of the evaluation sets and their use in the research questions.}
    \label{tab:dataset-rq-overview}
    \small
    \resizebox{\linewidth}{!}{%
    \begin{tabular}{p{0.24\linewidth}r p{0.23\linewidth} p{0.38\linewidth}}
        \toprule
        Evaluation set & \ No. of Commits & Sampling frame & Research question and purpose \\
        \midrule
        Main evaluation set & 7,522 & ApacheCM & \textbf{RQ1}: Overall effectiveness \\
        Entity annotation set & 100 & ApacheCM & \textbf{RQ2.1}: Core-entity identification \\
        Ablation subset & 1,000 & Main evaluation set & \textbf{RQ2.2}: View contribution analysis \\
        Human evaluation subset & 50 & Ablation subset & \textbf{RQ3}: Human evaluation \\
        \bottomrule
    \end{tabular}
    }
\end{table}

\subsection{Comparison Baselines}
\label{sec:baseline}

\rev{To comprehensively evaluate the effectiveness of CAMGen, we compare CAMGen with six existing CMG approaches and a diff-only LLM baseline. The existing approaches cover nearest-neighbor retrieval, retrieval-augmented neural generation, code-change pretraining, and LLM-based context augmentation. We select methods with public implementations and configuration details so that they can be evaluated on our dataset using a common evaluation pipeline. The diff-only LLM baseline additionally isolates the benefit of context augmentation under the same generator.}

\begin{itemize}
\item \citefix{\textbf{NNGen}~\cite{nngen}.} NNGen is a representative early CMG approach that relies on nearest-neighbor retrieval. Given a code diff, it searches a collection of historical commits for the most similar code change and directly uses the corresponding developer-written commit message as the output.

\item \citefix{\textbf{CoRec}~\cite{corec}.} CoRec combines retrieval with neural generation. It first retrieves information related to the current diff from historical data and then supplies the retrieved information to the generator. Unlike approaches that directly copy a retrieved message, CoRec generates a new message for the current change, enabling it to accommodate partial mismatches between the retrieved example and the target commit.

\item \citefix{\textbf{RACE}~\cite{race}.} RACE adopts a retrieval-augmented CMG strategy that uses historical code changes and their commit messages as additional generation context. Its central idea is to exploit expressions and change patterns from historical commits instead of relying solely on the local text of the current diff.

\item \citefix{\textbf{COME}~\cite{He2023COME}.} COME represents code changes through modification embeddings and uses these representations to generate or select candidate messages. By encoding each code change into a semantic representation suitable for CMG, COME addresses the limited ability of surface lexical similarity to capture change content.

\item \citefix{\textbf{CCT5}~\cite{Lin2023CCT5}.} CCT5 is a pretrained generation model designed for code changes. It learns correspondences between code edits and natural-language descriptions through pretraining tasks tailored to code changes. Unlike retrieval-based approaches, CCT5 primarily relies on the code-change knowledge acquired during pretraining to generate a commit message from the input diff.

\item \rev{\textbf{OMEGA}~\cite{Imani2025OMEGA}. OMEGA is an LLM-based CMG approach that augments code diffs with various software context, including method and class summaries, and fine-grained explanations generated by the Fine-grained Interactive Diff Explainer (FIDEX). The diff, context, and explanations are combined into a prompt for the LLM to generate a commit message. We use DeepSeek-V3.2 as its underlying LLM and extend its original Java-only implementation to C++ and Python following the same context-construction and diff-explanation workflow.}

\item \rev{\textbf{C3Gen}~\cite{Xiong2025C3GEN}. C3Gen, proposed in our previous work, retrieves repository context through code structure graphs and invocation and instantiation relationships of changed entities. We adapt its retrieval strategy by using core entities as anchors and applying role-aware context pruning, and incorporate the resulting approach as $\mathrm{View}_{\mathrm{Struct}}$ (Section~\ref{subsec:structural-view}). The $\mathrm{View}_{\mathrm{Struct}}$-only results in the ablation study and human evaluation represent the performance of this baseline with DeepSeek-V3.2 as the generator.}

\item \rev{\textbf{CoRaCMG}~\cite{CoRaCMG}. CoRaCMG, proposed in our previous work, is a retrieval-augmented CMG approach that selects historical diff--message pairs through hybrid retrieval to guide commit message generation. We retain its equally weighted fusion of normalized BM25 and Jina-based cosine similarity scores, restrict candidates to commits from the same repository whose timestamps are strictly earlier than that of the commit corresponding to the query code diff, and incorporate the resulting approach as $\mathrm{View}_{\mathrm{Exemplar}}$ (Section~\ref{subsec:historical-view}). The $\mathrm{View}_{\mathrm{Exemplar}}$-only results in the ablation study and human evaluation represent the performance of this baseline with DeepSeek-V3.2 as the generator.}

\item \rev{\textbf{Diff-only LLM}. This baseline directly prompts DeepSeek-V3.2 to generate a commit message from the raw code diff, without additional repository context or retrieved exemplars.}

\end{itemize}

To ensure a fair comparison and prevent test-data leakage throughout the CMG experiments, we first exclude the 7,522 commits in the RQ1 evaluation set from the full ApacheCM dataset. All training data required by the baselines are then sampled from the remaining records. We retrain CoRec, RACE, COME, and CCT5 following their public implementations and experimental settings. For NNGen, we reconstruct the historical candidate collection and retrieval index using only the remaining records. The retrieval corpus for the Historical Exemplar View of CAMGen is also constructed exclusively from the remaining records, excluding all 7,522 evaluation commits. For each target commit, retrieval is further restricted to earlier commits from the same repository, as described in Section~\ref{subsec:historical-view}. In each comparison experiment, all baselines and CAMGen share identical language partitions, evaluation samples, reference commit messages, preprocessing procedure, and evaluation pipeline.

\subsection{Evaluation Design}
This paper employs both automatic evaluation and human evaluation to assess the quality of LLM-generated commit messages. Automatic metrics are used to measure lexical, sequential, and semantic similarity between generated commit messages and reference commit messages, whereas human evaluation assesses content coverage, change rationale, change impact, and linguistic quality of generated commit messages from the developers' perspective.

\subsubsection{Automatic Evaluation}
\rev{Following prior work on CMG\citefix{~\cite{wu2025icl, Zeng2026}}, we evaluate the similarity between generated and reference commit messages using BLEU, ROUGE-L, METEOR, CIDEr, and SBERT Cosine. All methods are evaluated using these five metrics under the same preprocessing and evaluation pipeline. Higher values indicate better performance within each metric.}

\begin{itemize}
    \item \textbf{BLEU}~\cite{bleu}: Originally developed for machine translation evaluation, this metric computes n-gram precision between generated text and a reference. \rev{Following prior CMG studies~\cite{Xiong2025C3GEN,CoRaCMG}, we use the Google BLEU implementation provided by Hugging Face Evaluate in our evaluation.} In CMG, it reflects the extent to which a generated message reproduces key phrases and technical expressions from the reference commit message.
    \item \textbf{ROUGE-L}~\cite{rouge}: Based on the longest common subsequence between the generated and reference texts, this metric captures overlap in word order and sequential expression, thereby retaining more information about sentence structure than metrics based solely on independent n-grams.
    \item \textbf{METEOR}~\cite{meteor}: This metric computes the harmonic mean of unigram precision and recall and supports matches based on morphological variants and synonyms. Therefore, it evaluates whether the generated message conveys the primary meaning of the reference under a broader notion of lexical correspondence.
    \item \textbf{CIDEr}\citefix{~\cite{vedantam2015cider}}: This metric uses TF-IDF to weight n-grams, emphasizing words and phrases that are distinctive within the corpus. For commit messages that contain API names, module names, and project-specific terminologies, it complements BLEU by emphasizing matches that involve informative phrases.
    \item \textbf{SBERT Cosine}\citefix{~\cite{reimers2019sbert}}: This metric uses Sentence-BERT to encode the generated commit messages and the reference commit messages as semantic vectors and computes their cosine similarity. Rather than requiring identical surface wording, SBERT Cosine is used to measure proximity in a sentence-level semantic space.
\end{itemize}


\subsubsection{Human Evaluation}
\label{sec:human-eval}
This subsection first describes the human evaluation process and then defines the five evaluation dimensions.

\textbf{Human Evaluation Process.} 
To address RQ3, a human evaluation is conducted on the 50-commit evaluation set described in Section~\ref{sec:research-datasets}. For each commit, all three human evaluators receive the code diff, the reference commit message, and the candidate messages generated by the six existing CMG approaches (see Section~\ref{sec:baseline}), the diff-only baseline, and various view-derived context combinations of CAMGen. They also have access to repository snapshots from before and after the commit, which they can browse and search to inspect surrounding code and compare the two repository states when assessing change content, rationale, and impact. To ensure a blind evaluation, we randomly shuffle the candidate messages and remove all source-identifying information, including model names, method names, and context-configuration identifiers, to prevent bias.

We review substantial rating disagreements through discussion and retain a separate final rating from each evaluator. 
\rev{For each method or configuration, we average the three evaluator ratings within each commit and then average across the 50 commits. Following prior work on CMG evaluation~\cite{Zeng2026}, we report, for each metric, the mean of the three pairwise Spearman rank correlations~\cite{spearman1904} to assess the consistency of message rankings. We also report 95\% percentile bootstrap confidence intervals for Full versus OMEGA, Full versus Diff only, the three view-removal losses, and the two contrasts between these losses. For each comparison, we use the post-discussion ratings averaged within each commit and 50,000 paired resamples of the 50 commits, preserving the pairing across methods and configurations.}

\textbf{Human Evaluation Metrics.} The human evaluation adopts a 0-4 scale, with higher scores indicating better quality. \citefix{We adopt four dimensions, i.e., \textit{Rationality}, \textit{Comprehensiveness}, \textit{Conciseness}, and \textit{Expressiveness}~\cite{li2024only,li2026ccmg} from prior CMG studies.} 
Additionally, we introduce the fifth dimension, \textit{Change Impact Explanation}, to reflect the focus of this study by assessing whether a commit message explains how a change affects system functionality, program behavior, or module responsibilities.

\begin{itemize}
    \item \textbf{\textit{Comprehensiveness}}: This metric assesses whether a commit message accurately and sufficiently summarizes the primary change, including the key change targets, change actions, and affected scope. 
    \item \textbf{\textit{Rationality}}: This metric evaluates whether a commit message presents a logically coherent rationale, motivation, or intended objective that is consistent with the code change and thus explains why the change is made. We do not require a specific commit-type format because projects do not uniformly adopt the Conventional Commits specification.
    \item \textbf{\textit{Change Impact Explanation}}: This metric assesses whether a commit message explains how the change affects system functionality, program behavior, module responsibilities, user-visible capabilities, or maintenance activities. It evaluates whether the message transcends local edit operations to provide an accurate explanation of the change's impact, grounded in both the code and repository context.
    \item \textbf{\textit{Conciseness}}: This metric assesses whether a commit message retains the necessary change information while avoiding irrelevant, repetitive, or redundant content. \textit{Conciseness} does not simply require brevity; it requires a compact description focused on the primary change.
    \item \textbf{\textit{Expressiveness}}: This metric evaluates grammatical correctness, fluency, and clarity, with an emphasis on whether readers can understand the message naturally and accurately.
\end{itemize}

Human and automatic evaluations address complementary aspects: how closely a generated message resembles its reference and whether the message accurately and completely explains the code change.

\subsection{LLM Selection}
Table~\ref{tab:llm-selection} summarizes the LLMs used in our experiments and their respective tasks. For CMG, we use two open-source models (i.e., DeepSeek-V3.2 and DeepSeek-V4-Flash) and one closed-source model (i.e., GPT-5.4) to evaluate CAMGen across different generation models. For core entity identification in RQ2.1, we compare DeepSeek-V4-Flash and DeepSeek-V4-Pro, where DeepSeek-V4-Pro is used exclusively for this task. We set the temperature of all models to 0.0 to reduce generation randomness and keep decoding settings consistent across experiments.

\begin{table}[h]
    \centering
    \caption{LLMs selected for the experiments and supported tasks.}
    \label{tab:llm-selection}
    \resizebox{\linewidth}{!}{
    \begin{tabular}{lcccc}
        \toprule
        Model & Type & Release Date &
        \shortstack{CMG} &
        \shortstack{Core Entity Identification} \\
        \midrule
        DeepSeek-V3.2 & Open-source & December 2025 & $\checkmark$ & $\times$ \\
        DeepSeek-V4-Flash & Open-source & April 2026 & $\checkmark$ & $\checkmark$ \\
        DeepSeek-V4-Pro & Open-source & April 2026 & $\times$ & $\checkmark$ \\
        GPT-5.4 & Closed-source & March 2026 & $\checkmark$ & $\times$ \\
        \bottomrule
    \end{tabular}
    }
\end{table}
\section{Results and Analysis}
\label{sec:5}\label{5-results-and-analysis}

\subsection{Cross-Language Overall Performance of CAMGen (RQ1)}
RQ1 investigates the overall performance of CAMGen using three different LLMs as generators. We compare six existing CMG approaches and a diff-only LLM baseline on the same dataset for RQ1 using five automatic metrics. 
\rev{OMEGA, the Diff-only LLM baseline, and CAMGen with DeepSeek-V3.2 as the generator all use DeepSeek-V3.2, enabling a comparison under the same LLM for each programming language.}

\subsubsection{Results of RQ1}

\rev{Table~\ref{tab:rq1} presents the results for Java, C++, and Python, comparing NNGen, CoRec, RACE, COME, CCT5, OMEGA, and the diff-only LLM baseline with three CAMGen configurations using the full three-view context. Bold values indicate the highest score for each metric and programming language.}

\begin{table*}[t]
    \centering
    \caption{\rev{Performance of existing CMG methods and CAMGen on Java, C++, and Python.}}
    \label{tab:rq1}
    \resizebox{\textwidth}{!}{%
    \begin{tabular}{lccccccccccccccc}
        \toprule
        Method / LLM
        & \multicolumn{5}{c}{Java}
        & \multicolumn{5}{c}{C++}
        & \multicolumn{5}{c}{Python} \\
        \cmidrule(lr){2-6}
        \cmidrule(lr){7-11}
        \cmidrule(lr){12-16}
        & BLEU & ROUGE-L & METEOR & CIDEr & SBERT Cosine
        & BLEU & ROUGE-L & METEOR & CIDEr & SBERT Cosine
        & BLEU & ROUGE-L & METEOR & CIDEr & SBERT Cosine \\
        \midrule
        \multicolumn{16}{c}{\textit{Non-LLM CMG baselines}} \\
        \midrule
        NNGen~\cite{nngen}
        & 6.11 & 9.33 & 12.42 & 2.74 & 32.83
        & 6.57 & 12.32 & 14.01 & 3.72 & 35.50
        & 7.96 & 13.85 & 14.48 & 4.37 & 37.50 \\
        CoRec~\cite{corec}
        & 9.57 & 14.02 & 17.88 & 4.60 & 36.49
        & 7.63 & 14.82 & 15.69 & 3.67 & 36.75
        & 8.73 & 15.12 & 15.49 & 4.21 & 37.52 \\
        RACE~\cite{race}
        & 8.89 & 18.05 & 17.94 & 5.76 & 43.09
        & 7.89 & 12.13 & 8.62 & 2.68 & 40.06
        & 9.68 & 15.41 & 9.65 & 3.60 & 41.59 \\
        COME~\cite{He2023COME}
        & 11.79 & 19.43 & 19.00 & 9.22 & 46.35
        & 11.49 & 22.23 & 21.28 & 8.09 & 48.86
        & 10.42 & 21.99 & 14.55 & 8.30 & 48.39 \\
        CCT5~\cite{Lin2023CCT5}
        & 13.01 & 22.11 & 22.07 & 10.40 & 47.45
        & 12.21 & 23.91 & 22.55 & 8.82 & 50.49
        & 10.71 & 23.54 & 15.99 & 8.63 & 49.08 \\
        \midrule
        \multicolumn{16}{c}{\rev{\textit{LLM-based baselines (DeepSeek-V3.2)}}} \\
        \midrule
        \rev{Diff only}
        & \rev{9.71} & \rev{21.05} & \rev{18.83} & \rev{8.04} & \rev{53.15}
        & \rev{8.92} & \rev{22.72} & \rev{20.13} & \rev{7.34} & \rev{55.91}
        & \rev{10.53} & \rev{25.79} & \rev{22.13} & \rev{7.45} & \rev{57.36} \\
        \rev{OMEGA~\cite{Imani2025OMEGA}}
        & \rev{10.76} & \rev{22.75} & \rev{19.37} & \rev{10.06} & \rev{54.02}
        & \rev{9.62} & \rev{22.83} & \rev{19.46} & \rev{8.33} & \rev{55.95}
        & \rev{13.51} & \rev{\textbf{28.72}} & \rev{24.28} & \rev{11.57} & \rev{\textbf{59.19}} \\
        \midrule
        \multicolumn{16}{c}{\textit{CAMGen with different generator LLMs}} \\
        \midrule
        DeepSeek-V3.2
        & 11.42 & 22.98 & 20.11 & 10.50 & 53.55
        & 12.09 & 25.93 & 23.59 & 10.38 & 56.62
        & \textbf{14.01} & 28.63 & 24.10 & \textbf{11.87} & 58.43 \\
        DeepSeek-V4-Flash
        & \textbf{13.36} & \textbf{24.42} & \textbf{25.25} & \textbf{11.10} & \textbf{55.26}
        & \textbf{12.35} & \textbf{26.59} & \textbf{25.13} & \textbf{10.60} & \textbf{57.34}
        & 13.32 & 27.70 & \textbf{25.61} & 10.54 & 58.83 \\
        GPT-5.4
        & 12.75 & 22.99 & 23.02 & 10.54 & 53.94
        & 11.92 & 25.24 & 22.22 & 10.19 & 56.72
        & 12.61 & 26.36 & 22.75 & 10.14 & 57.88 \\
        \bottomrule
    \end{tabular}%
    }
\end{table*}

On Java, CAMGen driven by DeepSeek-V4-Flash achieves the best performance across all five metrics, with scores of 13.36 BLEU, 24.42 ROUGE-L, 25.25 METEOR, 11.10 CIDEr, and 55.26 SBERT Cosine. 
\rev{With DeepSeek-V3.2 empolyed as the generator, CAMGen exceeds OMEGA on BLEU, ROUGE-L, METEOR, and CIDEr (11.42 vs.\ 10.76, 22.98 vs.\ 22.75, 20.11 vs.\ 19.37, and 10.50 vs.\ 10.06), but obtains a lower SBERT Cosine score (53.55 vs.\ 54.02). CAMGen exceeds the diff-only LLM baseline on all five metrics.}

\rev{On C++, CAMGen with DeepSeek-V4-Flash achieves the highest BLEU, ROUGE-L, METEOR, CIDEr, and SBERT Cosine scores of 12.35, 26.59, 25.13, 10.60, and 57.34, respectively. Among the baselines, CCT5 leads on BLEU, ROUGE-L, METEOR, and CIDEr, whereas OMEGA has the highest SBERT Cosine score (55.95). With DeepSeek-V3.2 employed, CAMGen outperforms both OMEGA and the diff-only LLM baseline on all five metrics; its gains over OMEGA are 2.47, 3.10, 4.13, 2.05, and 0.67 points on BLEU, ROUGE-L, METEOR, CIDEr, and SBERT Cosine, respectively.}

\rev{On Python, CAMGen with DeepSeek-V3.2 achieves the highest BLEU and CIDEr scores (14.01 and 11.87), while CAMGen with DeepSeek-V4-Flash leads on METEOR (25.61). OMEGA obtains the highest ROUGE-L and SBERT Cosine scores (28.72 and 59.19), exceeding the best CAMGen scores on these metrics by 0.09 and 0.36 points, respectively. Under the same DeepSeek-V3.2 generator, CAMGen exceeds OMEGA on BLEU and CIDEr by 0.50 and 0.30 points but scores lower on ROUGE-L, METEOR, and SBERT Cosine. All three CAMGen configurations exceed the diff-only LLM baseline on all five metrics.}

\rev{Across all three languages, CAMGen with DeepSeek-V3.2 outperforms the same-model diff-only LLM baseline in all 15 language--metric comparisons. Its absolute gains range from 1.71 to 3.48 points on BLEU, 1.93 to 3.21 on ROUGE-L, 1.28 to 3.46 on METEOR, 2.46 to 4.42 on CIDEr, and 0.40 to 1.07 on SBERT Cosine. Against OMEGA under the same generator, CAMGen obtains higher scores in 11 of these 15 comparisons. The results therefore support a consistent benefit over direct diff-only generation and an advantage over OMEGA that varies by language and metric.}

\subsubsection{Analysis of RQ1}

\rev{\textbf{CAMGen outperforms the SOTA baselines on most metrics across the three languages.} The best-performing CAMGen configurations achieve the highest scores in 13 of the 15 comparisons across the three programming languages and five evaluation metrics. CAMGen with DeepSeek-V4-Flash leads on all five metrics for Java and C++, while the two DeepSeek generators obtain the highest BLEU, METEOR, and CIDEr scores for Python. On the remaining two metrics, Python ROUGE-L and SBERT Cosine, the best CAMGen scores are only 0.09 and 0.36 points lower than the strongest baseline, respectively. Moreover, with DeepSeek-V3.2 employed as the generator LLM, CAMGen outperforms direct diff-only baseline on all five metrics in each programming language. These results demonstrate the effectiveness of combining structural, functional, and historical context for CMG.}

\textbf{CAMGen substantially improves the semantic quality of commit messages.} The gains are particularly pronounced for METEOR and SBERT Cosine. METEOR accounts for recall and supports morphological and synonym matches, whereas SBERT Cosine directly measures sentence-level semantic similarity between the generated commit messages and reference commit messages. For Java, DeepSeek-V4-Flash achieves 25.25 METEOR and 55.26 SBERT Cosine, representing improvements of approximately 14.4\% and 16.5\% over CCT5, respectively. For C++, the two scores increase from 22.55 and 50.49 to 25.13 and 57.34. For Python, the best CAMGen scores reach 25.61 and 58.83, corresponding to improvements of approximately 60.2\% and 19.9\% over CCT5. The consistent gains across languages indicate that multi-view context helps LLMs capture change intent, functional scope, and impact rather than merely reproduce local terms from the reference commit message.

\textbf{CAMGen exhibits strong model adaptability and cross-language generalization across different LLMs and programming languages.} CAMGen driven by DeepSeek-V4-Flash achieves the best results on all five metrics for Java and C++, whereas DeepSeek-V3.2 obtains the highest \rev{CAMGen} BLEU, ROUGE-L, and CIDEr scores for Python and DeepSeek-V4-Flash obtains the highest \rev{CAMGen} METEOR and SBERT Cosine scores. GPT-5.4 also remains competitive across all three languages. The generator, therefore, influences which model achieves the best result for a particular metric, but the context jointly provided by the three views provides a common semantic input that benefits different models. Moreover, although Java, C++, and Python differ in their type systems, syntax, project organization, and commit-message conventions, \rev{CAMGen achieves the highest scores on most metrics across the three programming languages, with its best SBERT Cosine scores reaching} 55.26, 57.34, and 58.83, respectively. Within the scope of this evaluation, CAMGen, therefore, does not depend on a single LLM or the lexical characteristics of one language and demonstrates a context-modeling capability that generalizes across the three CMG language settings.

\begin{framed}
\begin{minipage}{0.94\linewidth}
\noindent \textbf{Key Findings of RQ1:}

\begin{itemize}
\item \rev{CAMGen outperforms the SOTA baselines on most evaluation metrics across Java, C++, and Python, achieving the highest scores in 13 of the 15 language--metric comparisons.}
\item CAMGen substantially improves METEOR and SBERT Cosine, indicating more accurate expression of change intent and functional semantics.
\item \rev{Although the best-performing model varies across metrics, CAMGen remains effective across all three LLMs and programming languages, demonstrating model adaptability and cross-language generalization.}
\end{itemize}
\end{minipage}
\end{framed}

\subsection{Reliability of Core Entity Identification in CAMGen (RQ2.1)}

RQ2.1 addresses the identification of core entities that serve as the foundation for context construction in CAMGen. Rather than comparing automatic metrics for CMG, this experiment compares model predictions with manually annotated core entities to determine whether the models identify the program entities around which the subsequent contexts should be constructed. Because a commit may modify multiple cooperating entities, we report both entity-level metrics and CmtAccuracy, which requires the complete core-entity set for a commit to be recovered correctly. We randomly sample 100 commits from ApacheCM~\cite{Xiong2025C3GEN} and manually annotate the core classes, functions, interfaces, and other program entities as the gold labels. DeepSeek-V4-Flash and DeepSeek-V4-Pro then identify core entities in the same commits, and we evaluate their predictions using Precision, Recall, F1, Accuracy, and commit-level CmtAccuracy.

\subsubsection{Results of RQ2.1}

Table~\ref{tab:rq2-1} reports the core-entity identification results of DeepSeek-V4-Flash and DeepSeek-V4-Pro for the 100 manually annotated commits. We compute Precision, Recall, F1, and Accuracy at the entity level, whereas CmtAccuracy requires the predicted core-entity set for a commit to match the gold labels exactly.

\begin{table}[h]
\centering
\caption{Core-entity identification results for RQ2.1.}
\label{tab:rq2-1}
\scalebox{0.7}{
\resizebox{\columnwidth}{!}{%
\begin{tabular}{lccccc}
\toprule
Model & Precision & Recall & F1 & Accuracy & CmtAccuracy \\
\midrule
DeepSeek-V4-Flash & 0.7862 & 0.8013 & 0.7937 & 0.7619 & 0.7300 \\
DeepSeek-V4-Pro & \textbf{0.8301} & \textbf{0.8141} & \textbf{0.8220} & \textbf{0.7985} & \textbf{0.7700} \\
\bottomrule
\end{tabular}%
}
}
\end{table}

DeepSeek-V4-Pro achieves the highest value for all five metrics, with Precision, Recall, F1, Accuracy, and CmtAccuracy scores of 0.8301, 0.8141, 0.8220, 0.7985, and 0.7700, respectively. The corresponding scores of DeepSeek-V4-Flash are 0.7862, 0.8013, 0.7937, 0.7619, and 0.7300. Relative to DeepSeek-V4-Flash, DeepSeek-V4-Pro improves the five metrics by 0.0439, 0.0128, 0.0283, 0.0366, and 0.0400, respectively, with the largest absolute difference observed for Precision.

At the entity level, both models achieve Recall above 0.80, with F1 scores of 0.7937 and 0.8220. At the commit level, DeepSeek-V4-Pro correctly recovers the complete core-entity set for 77 of the 100 commits, compared with 73 commits for DeepSeek-V4-Flash. The entity-level F1 and commit-level CmtAccuracy rankings are consistent, with DeepSeek-V4-Pro performing better at both evaluation granularities.

\subsubsection{Analysis of RQ2.1}

\textbf{Both LLMs identify core-entity sets accurately.} DeepSeek-V4-Flash obtains Precision, Recall, and F1 scores of 0.7862, 0.8013, and 0.7937, whereas DeepSeek-V4-Pro obtains 0.8301, 0.8141, and 0.8220. Recall exceeds 0.80 for both models, showing that both LLMs identify most manually annotated core entities, while their Precision scores indicate that they do not indiscriminately include every entity related to a modification. For CAMGen, this balance between precision and coverage is more important than maximizing either metric in isolation because $\mathrm{View}_{\mathrm{Struct}}$ and $\mathrm{View}_{\mathrm{Func}}$ use the core entities as shared anchors.

\textbf{DeepSeek-V4-Pro identifies core entities more accurately at both the entity and commit levels.} Compared with DeepSeek-V4-Flash, DeepSeek-V4-Pro increases Precision from 0.7862 to 0.8301 and Recall from 0.8013 to 0.8141. These improvements indicate both a higher proportion of correct predictions among the identified entities and greater coverage of the manually annotated core entities. F1 and Accuracy also increase from 0.7937 to 0.8220 and from 0.7619 to 0.7985, respectively. At the commit level, CmtAccuracy increases from 0.7300 to 0.7700: DeepSeek-V4-Pro predicts core-entity sets that exactly match the manual annotations for 77 of the 100 commits, compared with 73 for DeepSeek-V4-Flash. 
Compared with DeepSeek-V4-Flash, DeepSeek-V4-Pro therefore identifies individual core entities more accurately and correctly recovers the full set of core entities for a larger proportion of commits.

\begin{framed}
\begin{minipage}{0.94\linewidth}
\noindent \textbf{Key Findings of RQ2.1:}

\begin{itemize}
\item DeepSeek-V4-Flash and DeepSeek-V4-Pro both achieve strong core-entity identification performance against the manually annotated gold labels.
\item DeepSeek-V4-Pro outperforms DeepSeek-V4-Flash on all five metrics and retains this advantage at both the entity and commit levels.
\end{itemize}
\end{minipage}
\end{framed}

\subsection{Complementarity of Multi-View Repository Context (RQ2.2)}


\rev{RQ2.2 decomposes CAMGen into seven context-input configurations while holding the generator and samples constant. We use DeepSeek-V3.2 as the generator and randomly sample 1,000 commits from the combined Java, C++, and Python datset used in RQ1. The Diff-only baseline directly prompts the LLM to generate commit messages based solely on the code diff. 
\rev{The $\mathrm{View}_{\mathrm{Struct}}$-only and $\mathrm{View}_{\mathrm{Exemplar}}$-only configurations use the improved C3Gen and CoRaCMG baselines defined in Section~\ref{sec:baseline}}. The remaining configurations examine the contributions of structural and functional context individually and in combination with historical exemplars.}

\subsubsection{Results of RQ2.2}

Table~\ref{tab:rq2-2} reports the ablation results obtained by DeepSeek-V3.2 for 1,000 randomly sampled commits. All configurations include the raw code diff. Diff only adds no repository context. The single-view configurations add context provided by $\mathrm{View}_{\mathrm{Struct}}$, $\mathrm{View}_{\mathrm{Func}}$, or $\mathrm{View}_{\mathrm{Exemplar}}$, whereas the other configurations combine context provided by two or all three views.



\begin{table}[t]
\centering
\caption{\rev{Ablation results for the Diff-only baseline and seven CAMGen configurations (RQ2.2).}}
\label{tab:rq2-2}
\scalebox{0.8}{
\resizebox{\columnwidth}{!}{%
\begin{tabular}{lrrrrr}
\toprule
\rev{Method/Configuration} & BLEU & ROUGE-L & METEOR & CIDEr & SBERT Cosine \\
\midrule
\multicolumn{6}{c}{\rev{\textit{Baseline}}} \\
\midrule
Diff-only & 9.92 & 23.10 & 20.41 & 8.07 & 56.04 \\
\midrule
\multicolumn{6}{c}{\rev{\textit{CAMGen configurations}}} \\
\midrule
$\mathrm{View}_{\mathrm{Struct}}$ only & 11.01 & 24.07 & 20.73 & 9.67 & 55.72 \\
$\mathrm{View}_{\mathrm{Func}}$ only & 10.96 & 24.02 & 21.27 & 9.59 & 56.11 \\
$\mathrm{View}_{\mathrm{Exemplar}}$ only & 12.30 & 25.35 & 21.36 & 11.31 & 56.21 \\
$\mathrm{View}_{\mathrm{Struct}} + \mathrm{View}_{\mathrm{Func}}$ & 11.06 & 24.40 & 21.53 & 9.62 & 55.98 \\
$\mathrm{View}_{\mathrm{Struct}} + \mathrm{View}_{\mathrm{Exemplar}}$ & 12.86 & 25.76 & 22.98 & 11.67 & 56.82 \\
$\mathrm{View}_{\mathrm{Func}} + \mathrm{View}_{\mathrm{Exemplar}}$ & 12.66 & 25.94 & 22.92 & 11.28 & 56.74 \\
\midrule
CAMGen (Full) & \textbf{13.05} & \textbf{26.36} & \textbf{23.35}
        & \textbf{11.73} & \textbf{57.10} \\
\bottomrule
\end{tabular}%
}
}
\end{table}

Compared with the diff-only baseline, all three single-view context configurations achieve higher BLEU, ROUGE-L, METEOR, and CIDEr scores. Adding the structural-dependency context provided by $\mathrm{View}_{\mathrm{Struct}}$ improves these four metrics by approximately 11.0\%, 4.2\%, 1.6\%, and 19.8\%, respectively. Adding the functional-semantic context provided by $\mathrm{View}_{\mathrm{Func}}$ improves all five metrics, with relative gains of approximately 10.5\%, 4.0\%, 4.2\%, 18.8\%, and 0.1\% in BLEU, ROUGE-L, METEOR, CIDEr, and SBERT Cosine, respectively. Adding the historical-exemplar context provided by $\mathrm{View}_{\mathrm{Exemplar}}$ achieves the highest single-view scores on all five metrics, with relative improvements of approximately 24.0\%, 9.7\%, 4.7\%, 40.1\%, and 0.3\% over the diff-only baseline in BLEU, ROUGE-L, METEOR, CIDEr, and SBERT Cosine, respectively.

The two-view configurations exhibit different strengths across metrics. $\mathrm{View}_{\mathrm{Struct}} + \mathrm{View}_{\mathrm{Func}}$ obtains 11.06 BLEU, 24.40 ROUGE-L, 21.53 METEOR, 9.62 CIDEr, and 55.98 SBERT Cosine, whereas $\mathrm{View}_{\mathrm{Struct}} + \mathrm{View}_{\mathrm{Exemplar}}$ obtains 12.86, 25.76, 22.98, 11.67, and 56.82, respectively. $\mathrm{View}_{\mathrm{Func}} + \mathrm{View}_{\mathrm{Exemplar}}$ obtains 12.66, 25.94, 22.92, 11.28, and 56.74 on these five metrics, respectively. Among the two-view configurations, $\mathrm{View}_{\mathrm{Struct}} + \mathrm{View}_{\mathrm{Exemplar}}$ achieves the highest BLEU, METEOR, CIDEr, and SBERT Cosine scores, whereas $\mathrm{View}_{\mathrm{Func}} + \mathrm{View}_{\mathrm{Exemplar}}$ achieves the highest ROUGE-L score. Relative to the diff-only baseline, $\mathrm{View}_{\mathrm{Func}} + \mathrm{View}_{\mathrm{Exemplar}}$ improves BLEU, ROUGE-L, METEOR, CIDEr, and SBERT Cosine by approximately 27.6\%, 12.3\%, 12.3\%, 39.8\%, and 1.2\%, respectively.

The complete three-view context configuration of CAMGen achieves the highest scores among all seven configurations on all five metrics, obtaining 13.05 BLEU, 26.36 ROUGE-L, 23.35 METEOR, 11.73 CIDEr, and 57.10 SBERT Cosine. Compared with the diff-only baseline, these scores represent relative improvements of approximately 31.6\%, 14.1\%, 14.4\%, 45.4\%, and 1.9\%, respectively. Compared with $\mathrm{View}_{\mathrm{Struct}} + \mathrm{View}_{\mathrm{Exemplar}}$, which achieves the highest scores among the ablated configurations on four of the five metrics, the complete three-view context configuration of CAMGen further improves BLEU, ROUGE-L, METEOR, CIDEr, and SBERT Cosine by approximately 1.5\%, 2.3\%, 1.6\%, 0.5\%, and 0.5\%, respectively. It also outperforms the remaining ablated configurations on every metric.

\subsubsection{Analysis of RQ2.2}


\textbf{Incorporating context from individual views consistently improves automatic evaluation metrics over the diff-only baseline, with historical exemplars yielding the strongest single-view performance.} Specifically, adding context from ViewStruct, ViewFunc, or ViewExemplar enhances BLEU, ROUGE-L, METEOR, and CIDEr scores. The Structural Dependency View and Functional Semantic View enrich the LLM's input with repository-level code context surrounding core entities, whereas the Historical Exemplar View supplies paired examples of code diffs and developer-written commit messages. This additional context enables the LLM to generate commit messages that are more semantically accurate and stylistically consistent. Notably, ViewExemplar achieves the highest single-view scores across all five metrics. This advantage likely stems from historical exemplars providing project-specific terminology and writing styles for changes similar to the current code diff, thereby allowing the LLM to produce messages that align more closely with project conventions.

\textbf{Repository code context further improves semantic alignment when historical exemplars are already available.} Adding either $\mathrm{View}_{\mathrm{Struct}}$ or $\mathrm{View}_{\mathrm{Func}}$ to $\mathrm{View}_{\mathrm{Exemplar}}$ improves both METEOR and SBERT Cosine. Historical exemplars provide examples of how developers describe similar changes, while repository code supplies context for interpreting the current diff. Specifically, $\mathrm{View}_{\mathrm{Struct}}$ exposes dependencies around the core entities. $\mathrm{View}_{\mathrm{Func}}$ adds code from candidate functional units containing these entities, helping the LLM examine the responsibilities of the changed code and the implications of the edits. Combining this information with historical examples allows the LLM to describe the current change using project-specific terminology, offering an explanation for the observed gains in semantic alignment with the reference descriptions.

\textbf{The complete three-view context configuration of CAMGen achieves the best performance across all metrics, and removing any view consistently degrades performance.} This indicates that none of the three views is redundant: each contributes complementary information that the LLM relies on to generate more accurate and informative commit messages. $\mathrm{View}_{\mathrm{Struct}}$ and $\mathrm{View}_{\mathrm{Func}}$ primarily enrich the representation of the code change itself, exposing the structural dependencies and functional semantics around the modified entities. In contrast, $\mathrm{View}_{\mathrm{Exemplar}}$ complements this information by providing examples of similar diffs and human-written commit messages, which help the LLM align with project-specific terminology and expression styles. Together, the three views enable the LLM to reason about what has changed, why the change matters, and how it should be described, which helps the model generate commit messages that are more accurate, informative, and consistent with project conventions.

\begin{framed}
\begin{minipage}{0.94\linewidth}
\noindent \textbf{Key Findings of RQ2.2:}

\begin{itemize}
\item The context provided by each view improves BLEU, ROUGE-L, METEOR, and CIDEr over the diff-only baseline, with historical exemplars achieving the highest single-view scores across all metrics.
\item The complete three-view context configuration \rev{achieves the best performance among the seven CAMGen configurations and outperforms the Diff-only baseline on all automatic evaluation metrics}. Removing any of the three views reduces all five scores, supporting their complementary contributions.
\end{itemize}
\end{minipage}
\end{framed}

\subsection{Human Evaluation of Commit Message Quality (RQ3)}

RQ3 uses human evaluation to assess aspects of generated commit message quality that automatic metrics cannot fully capture. Automatic metrics compare a generated message with the reference commit message, whereas human evaluation further determines whether the message accurately describes the code change target, rationale, and impact. We sample 50 commits from the RQ2.2 ablation subset and provide three human evaluators with the code diffs, reference commit messages, anonymized generated messages, and access to browsable and searchable repository snapshots from before and after each commit. Under the same evaluation protocol, we compare six baselines, the diff-only baseline, and seven view-context combinations. We report mean final ratings across the three evaluators and 50 commits, together with 95\% confidence intervals for the comparison between CAMGen(Full) and the diff-only baseline, the three view-removal losses, and the two contrasts between these losses.

\subsubsection{Results of RQ3}

Table~\ref{tab:rq3-human} presents the mean human ratings for the six existing CMG approaches and seven context configurations. All dimensions use a 0--4 scale, with higher scores indicating better quality. Bold values identify the highest mean rating for each dimension.

\begin{table*}[t]
\centering
\caption{Mean human ratings for CAMGen and the baselines on 50 commits (RQ3).}
\label{tab:rq3-human}
\resizebox{\textwidth}{!}{%
\begin{tabular}{lrrrrr}
\toprule
Method or Configuration & \textit{Comprehensiveness} & \textit{Rationality} & \textit{Change Impact Explanation} & \textit{Conciseness} & \textit{Expressiveness} \\
\midrule
NNGen~\cite{nngen} & 0.39 & 0.23 & 0.21 & 0.87 & 1.55 \\
CoRec~\cite{corec} & 0.37 & 0.17 & 0.13 & 0.86 & 1.28 \\
RACE~\cite{race} & 1.42 & 0.65 & 0.66 & 1.90 & 2.01 \\
COME~\cite{He2023COME} & 1.63 & 0.71 & 0.80 & 2.24 & 2.42 \\
CCT5~\cite{Lin2023CCT5} & 1.90 & 1.01 & 1.15 & 2.53 & 2.63 \\
\rev{OMEGA~\cite{Imani2025OMEGA}} & \rev{3.15} & \rev{2.87} & \rev{2.95} & \rev{3.19} & \rev{3.74} \\
Diff only & 2.65 & 2.30 & 2.48 & 2.82 & 3.68 \\

\midrule\midrule
$\mathrm{View}_{\mathrm{Struct}}$ only & 2.78 & 2.41 & 2.61 & 2.87 & \textbf{3.75} \\
$\mathrm{View}_{\mathrm{Func}}$ only & 3.05 & 2.73 & 2.83 & 3.03 & 3.72 \\
$\mathrm{View}_{\mathrm{Exemplar}}$ only & 2.96 & 2.59 & 2.74 & 2.96 & 3.70 \\
$\mathrm{View}_{\mathrm{Struct}}$ + $\mathrm{View}_{\mathrm{Func}}$ & 3.17 & 2.85 & 2.92 & 3.13 & \textbf{3.75} \\
$\mathrm{View}_{\mathrm{Struct}}$ + $\mathrm{View}_{\mathrm{Exemplar}}$ & 3.01 & 2.69 & 2.79 & 3.01 & 3.71 \\
$\mathrm{View}_{\mathrm{Func}}$ + $\mathrm{View}_{\mathrm{Exemplar}}$ & 3.14 & 2.88 & 2.94 & 3.15 & 3.69 \\
CAMGen (Full) & \textbf{3.34} & \textbf{3.05} & \textbf{3.08} & \textbf{3.29} & 3.73 \\
\bottomrule
\end{tabular}%
}
\end{table*}


\rev{Among the seven existing SOTA CMG baselines, OMEGA achieves the highest ratings on all five dimensions, with scores of 3.15, 2.87, 2.95, 3.19, and 3.74 for \textit{Comprehensiveness, Rationality, Change Impact Explanation, Conciseness, and Expressiveness}, respectively. CCT5 achieves the highest ratings among the five non-LLM baselines.}

The diff-only baseline obtains scores of 2.65, 2.30, 2.48, 2.82, and 3.68 on the five dimensions, respectively. Each single-view configuration of CAMGen improves the first four dimensions over this baseline. Among the single-view configurations of CAMGen, $\mathrm{View}_{\mathrm{Func}}$ achieves the highest \textit{Comprehensiveness}, \textit{Rationality}, \textit{Change Impact Explanation}, and \textit{Conciseness} scores, reaching 3.05, 2.73, 2.83, and 3.03, respectively. The highest reported \textit{Expressiveness} score among all configurations is 3.75, shared by $\mathrm{View}_{\mathrm{Struct}}$ alone and $\mathrm{View}_{\mathrm{Struct}}+\mathrm{View}_{\mathrm{Func}}$.

Among the two-view configurations, $\mathrm{View}_{\mathrm{Struct}}+\mathrm{View}_{\mathrm{Func}}$ achieves the highest \textit{Comprehensiveness} score of 3.17, whereas $\mathrm{View}_{\mathrm{Func}}+\mathrm{View}_{\mathrm{Exemplar}}$ achieves the highest \textit{Rationality}, \textit{Change Impact Explanation}, and \textit{Conciseness} scores of 2.88, 2.94, and 3.15. The complete three-view context configuration of CAMGen exceeds all single-view and two-view configurations on these four dimensions, reaching 3.34, 3.05, 3.08, and 3.29, respectively. \rev{Compared with OMEGA, the complete configuration of CAMGen improves mean \textit{Comprehensiveness}, \textit{Rationality}, \textit{Change Impact Explanation}, and \textit{Conciseness} by approximately 0.19, 0.18, 0.13, and 0.10 points, respectively. Its \textit{Expressiveness} score is 3.73, compared with 3.74 for OMEGA.}

Table~\ref{tab:rq3-ci} presents the 95\% confidence intervals for the paired comparisons. Compared with the diff-only baseline, the complete three-view context configuration of CAMGen improves \textit{Comprehensiveness, Rationality, Change Impact Explanation, Conciseness, and Expressiveness} by approximately 0.69, 0.75, 0.60, 0.47, and 0.05 points, respectively, with all five confidence intervals lying above zero. \rev{Compared with OMEGA, the full CAMGen configuration also achieves improvements whose confidence intervals lie entirely above zero for the first four dimensions.} For the first four dimensions, the confidence intervals for all three view-removal losses ($\Delta_{\mathrm{Struct}}$, $\Delta_{\mathrm{Func}}$, and $\Delta_{\mathrm{Exemplar}}$) lie above zero. The intervals for $\Delta_{\mathrm{Func}}-\Delta_{\mathrm{Struct}}$ and $\Delta_{\mathrm{Func}}-\Delta_{\mathrm{Exemplar}}$ also lie above zero on these four dimensions.

\begin{table*}[t]
\centering
\caption{95\% percentile bootstrap confidence intervals for paired mean rating differences in RQ3.}
\label{tab:rq3-ci}
\resizebox{\textwidth}{!}{%
\begin{tabular}{lccccc}
\toprule
Comparison & \shortstack{Comprehen-\\siveness} & Rationality & \shortstack{Change Impact\\Explanation} & Conciseness & Expressiveness \\
\midrule
\rev{Full $-$ OMEGA}
& \rev{$[0.107,\,0.287]$} & \rev{$[0.100,\,0.260]$} & \rev{$[0.047,\,0.220]$} & \rev{$[0.013,\,0.180]$} & \rev{$[-0.047,\,0.033]$} \\

Full $-$ Diff only
& $[0.607,\,0.767]$ & $[0.667,\,0.827]$ & $[0.513,\,0.687]$ & $[0.380,\,0.560]$ & $[0.013,\,0.100]$ \\

$\Delta_{\mathrm{Struct}}$
& $[0.133,\,0.267]$ & $[0.107,\,0.227]$ & $[0.073,\,0.213]$ & $[0.080,\,0.213]$ & $[0.013,\,0.093]$ \\

$\Delta_{\mathrm{Func}}$
& $[0.247,\,0.407]$ & $[0.273,\,0.433]$ & $[0.200,\,0.380]$ & $[0.200,\,0.353]$ & $[-0.007,\,0.060]$ \\

$\Delta_{\mathrm{Exemplar}}$
& $[0.100,\,0.253]$ & $[0.133,\,0.267]$ & $[0.100,\,0.227]$ & $[0.100,\,0.227]$ & $[-0.040,\,0.013]$ \\

\midrule
$\Delta_{\mathrm{Func}}-\Delta_{\mathrm{Struct}}$
& $[0.073,\,0.187]$ & $[0.113,\,0.267]$ & $[0.093,\,0.207]$ & $[0.080,\,0.193]$ & $[-0.067,\,0.007]$ \\

$\Delta_{\mathrm{Func}}-\Delta_{\mathrm{Exemplar}}$
& $[0.100,\,0.213]$ & $[0.087,\,0.220]$ & $[0.073,\,0.187]$ & $[0.060,\,0.173]$ & $[-0.007,\,0.080]$ \\
\bottomrule
\end{tabular}%
}
\par\smallskip
\begin{minipage}{\textwidth}
\footnotesize
$\Delta_v$ denotes the mean rating of Full minus that of the
configuration with view $v$ removed.
$\Delta_{\mathrm{Func}}-\Delta_{\mathrm{Struct}}$ is the rating
loss from removing the functional view minus the loss from
removing the structural view;
$\Delta_{\mathrm{Func}}-\Delta_{\mathrm{Exemplar}}$ is the rating
loss from removing the functional view minus the loss from
removing the historical exemplar view.
\end{minipage}
\end{table*}

Table~\ref{tab:rq3-agreement} reports the mean pairwise Spearman correlations calculated from the three sets of individual ratings retained after discussion. The values are 0.6134, 0.6525, 0.6572, 0.6194, and 0.6505 for \textit{Comprehensiveness}, \textit{Rationality}, \textit{Change Impact Explanation}, \textit{Conciseness}, and \textit{Expressiveness}, respectively. These correlations describe agreement in the post-discussion rankings.

\begin{table}[t]
\centering
\caption{\rev{Mean pairwise Spearman correlation among the three human evaluators.}}
\label{tab:rq3-agreement}
\scalebox{0.8}{
\begin{tabular}{lr}
\toprule
Human Evaluation Metric & Pairwise Spearman Mean \\
\midrule
\textit{Comprehensiveness} & \rev{0.6134} \\
\textit{Rationality} & \rev{0.6525} \\
\textit{Change Impact Explanation} & \rev{0.6572} \\
\textit{Conciseness} & \rev{0.6194} \\
\textit{Expressiveness} & \rev{0.6505} \\
\bottomrule
\end{tabular}
}
\end{table}

\subsubsection{Analysis of RQ3}
\textbf{Richer context helps CAMGen generate more informative and focused commit messages.} The concurrent improvements in \textit{Conciseness} and \textit{Comprehensiveness} indicate that the additional context enables the LLM to retain essential change information while filtering out irrelevant or repetitive content. \textit{Expressiveness}, by contrast, varies little across the seven CAMGen configurations.
Since this metric primarily assesses grammatical correctness, fluency, and clarity, the similar scores across configurations are expected, as all messages are generated by the same underlying LLM.  These results suggest that CAMGen primarily enhances the LLM's understanding of code diffs, enabling it to generate commit messages that more fully describe the main change, explain its rationale and impact, and remain focused on relevant information.

\textbf{The complete three-view context configuration of CAMGen achieves the highest human ratings for comprehensiveness, rationality, change impact explanation, and conciseness.} 
\rev{CAMGen (Full) receives higher mean ratings than all five non-LLM SOTA baselines across all five evaluation dimensions. Compared with the LLM-based SOTA baseline OMEGA, CAMGen (Full) also achieves higher mean ratings on the first four dimensions.} The observed rating differences indicate that the combined context improves how fully the messages describe the primary change, explain its rationale, and identify its impact while keeping the description focused. This pattern supports the goal of CAMGen: connecting local edits to the functionality they affect so that the generated message conveys the main change and its implications.

\textbf{Removing the Functional Semantic View causes the largest drops in \textit{Comprehensiveness, Rationality, Change Impact Explanation}, and \textit{Conciseness}.} The confidence intervals for both contrasts between view-removal losses in Table~\ref{tab:rq3-ci} lie above zero on these dimensions. \rev{Removing $\mathrm{View}_{\mathrm{Func}}$ leaves $\mathrm{View}_{\mathrm{Struct}}$+$\mathrm{View}_{\mathrm{Exemplar}}$, the C3Gen+CoRaCMG configuration. The positive paired intervals for $\Delta_{\mathrm{Func}}$ on these four dimensions therefore support the additional contribution of functional context beyond structural retrieval and historical exemplars.} $\mathrm{View}_{\mathrm{Func}}$ also achieves the highest single-view ratings on these four dimensions. 
\rev{Together, these results support the value of the context constructed by $\mathrm{View}_{\mathrm{Func}}$ for explaining code changes. The selected clusters supply related code around the core entities, giving the LLM additional information for interpreting the scope and behavioral impact of the code diff.}
Notably, while historical exemplars dominate the auto evaluation metrics in RQ2.2, functional context proves most valuable for the human-rated dimensions that assess how well a commit message explains the code change.

\textbf{Structural dependencies and historical exemplars each provide complementary gains beyond functional context.} The positive confidence intervals for $\Delta_{\mathrm{Struct}}$ and $\Delta_{\mathrm{Exemplar}}$ show that removing either view from Full reduces \textit{Comprehensiveness}, \textit{Rationality}, \textit{Change Impact Explanation}, and \textit{Conciseness}. These improvements reflect the distinct information each view supplies to the LLM: structural context exposes explicit call, inheritance, and instantiation relationships, whereas historical exemplars offer project-specific terminology and writing conventions for describing similar code diffs. By providing the LLM with these complementary sources of information, CAMGen enables the generation of commit messages that offer more complete and focused explanations of change rationale and impact.

\begin{framed}
\begin{minipage}{0.94\linewidth}
\noindent \textbf{Key Findings of RQ3:}

\begin{itemize}
\item The complete three-view context configuration achieves the highest mean human ratings for \textit{Comprehensiveness}, \textit{Rationality}, \textit{Change Impact Explanation}, and \textit{Conciseness}, with gains over the Diff-only baseline \rev{and OMEGA} supported by confidence intervals above zero on all four dimensions.
\item The Functional Semantic View achieves the highest single-view ratings on these four dimensions. Its removal from the full configuration also leads to the largest mean rating drops, a conclusion further supported by the fact that the confidence intervals for both contrasts between view-removal losses lie above zero.
\item Structural dependencies and historical exemplars provide further improvements on all four dimensions when added to configurations containing functional context, supporting more complete and focused explanations of code changes.
\end{itemize}
\end{minipage}
\end{framed}

\subsection{Qualitative Analysis of Low-Scoring Cases (RQ4)}

\subsubsection{Results of RQ4}
In this RQ, we aim to further understand the causes of cases in which CAMGen-generated commit messages receive low scores on automatic metrics. We manually inspect the underperforming cases from RQ1 by examining the code diff, the reference commit message, the context constructed or retrieved by CAMGen, and the generated message. The analysis identifies three main categories of causes: \textbf{Core-Entity Misidentification}, \textbf{Assessment Mismatch}, and \textbf{Incorrect Generation}. Assessment Mismatch further includes two subcategories: \textbf{Low-quality Reference} and \textbf{Semantically Similar Messages}. We discuss these categories of causes and representative cases below.

\paragraph{\textbf{Category I: Core-Entity Misidentification.}}

This category refers to cases in which CAMGen identifies entities that do not represent the primary intent of the commit while failing to identify the entity that carries the central change semantics. Because the identified entities serve as anchors for subsequent context construction, an incorrect anchor can affect the structural and functional context supplied to the generator and lead to an inaccurate commit message.

\begin{figure}[b]
    \centering
    \includegraphics[width=0.8\linewidth]{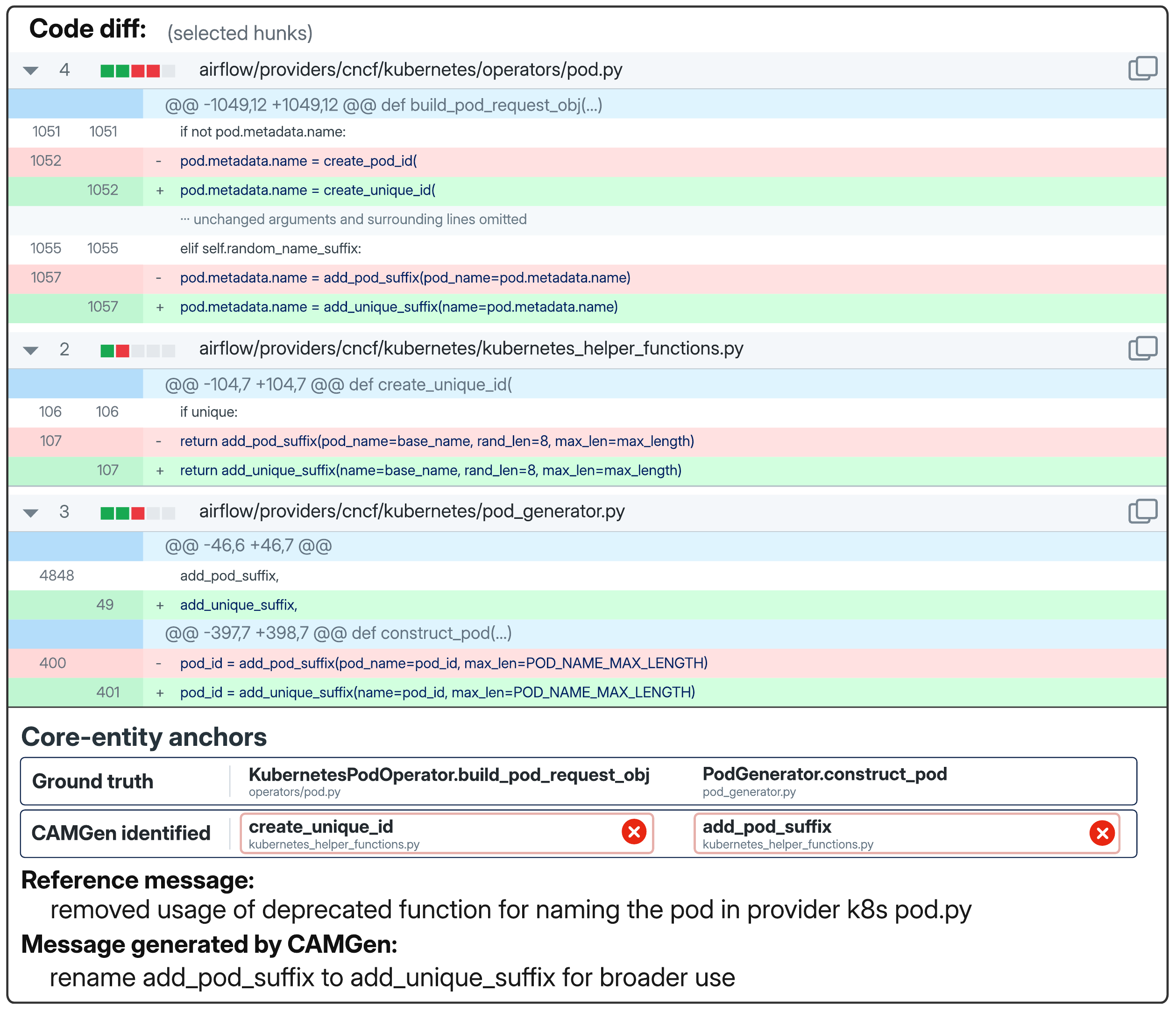}
    \caption{An example of Core-Entity Misidentification.}
    \label{fig:rq4-core-entity}
\end{figure}

As shown in Figure~\ref{fig:rq4-core-entity}, the primary purpose of the Airflow commit{\footnote{\citefix{\url{https://github.com/apache/airflow/commit/a19a9cb52388118e5fc735a25cc42229576482ad}}}} is to remove the use of deprecated pod-naming functions in the Kubernetes pod operator. However, CAMGen anchors on a helper function whose internal call is modified as a secondary consequence of the migration, while failing to identify the main change in \texttt{operators/pod.py}. As a result, the generated messages focus on renaming helper functions and omit the removal of deprecated naming usage in the pod operator. This example shows how an incorrect core-entity anchor can propagate to subsequent context construction and cause the generated message to miss the primary functional intent of the commit.

\begin{figure}[b]
    \centering
    \includegraphics[width=0.8\linewidth]{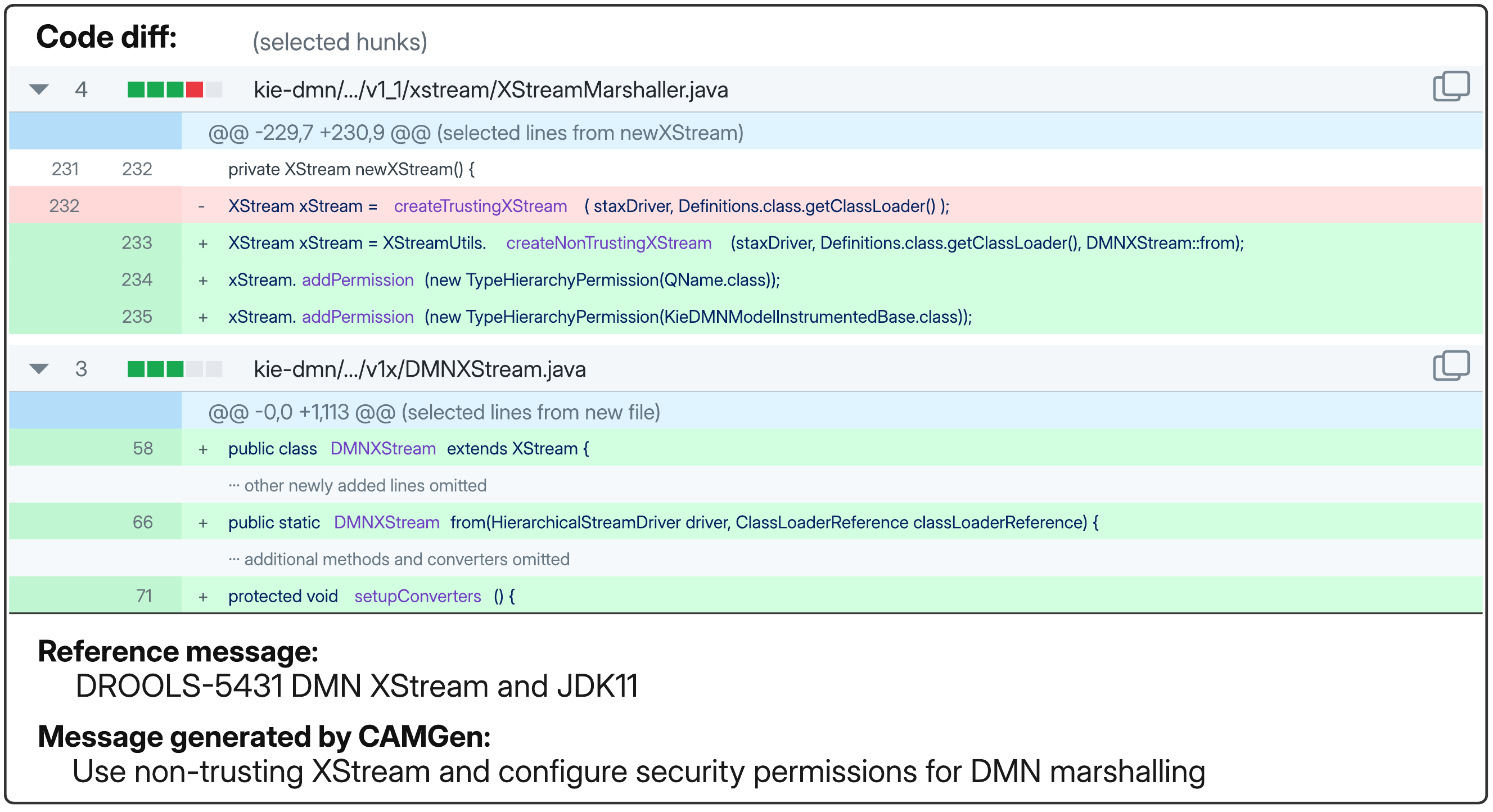}
    \caption{An example of a low-quality reference commit message.}
    \label{fig:rq4-low-quality-reference}
\end{figure}

\paragraph{\textbf{Category II: Assessment Mismatch.}}
Unlike the other two categories, Assessment Mismatch does not necessarily indicate an error in message generation. Instead, the generated message may be semantically valid, whereas its low automatic metric score results from a mismatch between the generated message and the reference commit message. We identify two subcategories in this group.

\textbf{(1) Low-quality Reference.} This subcategory refers to cases in which the reference commit message is too general or contains insufficient information about the actual code change. As shown in Figure~\ref{fig:rq4-low-quality-reference}, the reference commit message ``\textit{DROOLS-5431 DMN XStream and JDK11}''\citefix{\footnote{\citefix{\url{https://github.com/apache/incubator-kie/commit/b2db6fd531aa17074fb57c2e654a376df09ceeaf}}}} provides little information about the code diff. In contrast, the generated messages describe the replacement of the trusting XStream configuration and the addition of explicit type permissions. Although these messages provide substantially more information about the code diff, their automatic metric scores remain low because the reference commit message does not contain the corresponding content. This example indicates that low automatic metric scores may reflect the limited descriptive quality of the reference commit message rather than poor generation quality.

\begin{figure}[h]
    \centering
    \includegraphics[width=0.8\linewidth]{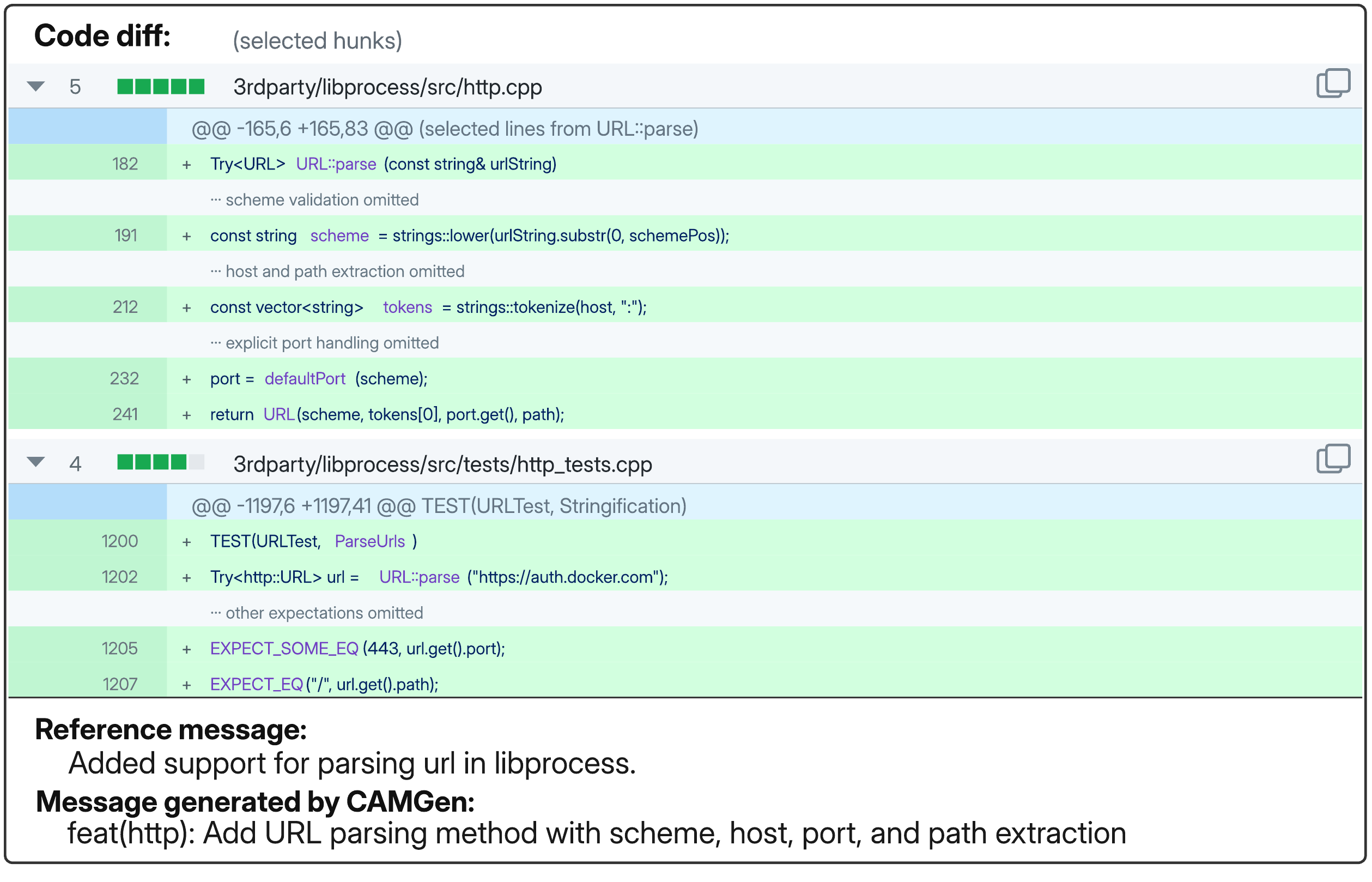}
    \caption{An example of Semantically Similar Messages.}
    \label{fig:rq4-semantically-similar}
\end{figure}

\textbf{(2) Semantically Similar Messages.} This subcategory refers to cases in which the generated message and the reference commit message correctly convey the same change intent but use different words or sentence structures. As shown in Figure~\ref{fig:rq4-semantically-similar}, the reference commit message ``\textit{Added support for parsing url in libprocess}''\citefix{\footnote{\citefix{\url{https://github.com/apache/mesos/commit/389e0333cb24727a21ec1ec9e30ddf230a1aa489}}}} and the generated message ``\textit{feat(http): Add URL parsing method with scheme, host, port, and path extraction}'' both describe the addition of URL parsing support. However, the two messages have limited lexical overlap because the generated message uses different terminologies and provides a more detailed description. Consequently, the generated message receives low lexical metric scores despite being semantically consistent with the reference. This example demonstrates that a single reference commit message may not capture all valid ways of describing the same code change.\citefix{ Similar observations are reported in prior CMG evaluation studies~\cite{wu2025icl,Zeng2026}.}

\paragraph{\textbf{Category III: Incorrect Generation.}}
This category refers to cases in which the LLM generates an incorrect commit message despite relevant evidence in the supplied code diff and repository context. The generated commit message may mischaracterize the defect being fixed or emphasize secondary edits while failing to convey the primary purpose of the commit. These errors suggest difficulties in interpreting change semantics and distinguishing the main change from accompanying modifications. Similar failures are reported by Wu et al.~\cite{wu2025icl}, who identified model fallacy as a category of incorrect commit message generation attributed to limitations of the underlying LLM. 
CAMGen addresses this by providing structural, functional, and historical context to help the LLM better understand the code diff, enabling the LLM to generate more accurate commit messages that correctly capture the primary code change and its functional impact.

\section{Threats to Validity}
\label{sec:6}\label{6-threats-to-validity}
Following the empirical research guidelines of \citefix{Wohlin et al.~\cite{wohlin2012experimentation}}, we discuss the factors that may affect our findings in terms of internal, external, construct, and conclusion validity. For each threat, we describe its potential effect and the mitigation measures adopted in the experiments or planned for future improvements.

\subsection{Internal Validity}
Internal validity concerns potential biases or errors introduced by the experimental design and execution.

LLM configurations and prompt design may introduce experimental bias. Model versions, prompt templates, and decoding parameters can affect the generated messages. In RQ1, all three LLMs use the same CAMGen input structure and generation requirements, with the temperature fixed at 0.0. 
\rev{OMEGA and the Diff-only LLM baseline use DeepSeek-V3.2, allowing comparison with CAMGen under the same generation model on all 7,522 commits. We implement OMEGA's context-construction and diff-explanation procedures for C++ and Python to extend its original Java support. We retrain CoRec, RACE, COME, and CCT5 and rebuild NNGen's retrieval index following their public implementations and settings. In RQ2.2, we keep the generator, samples, generation instructions, and evaluation scripts fixed. The $\mathrm{View}_{\mathrm{Exemplar}}$ of CAMGen uses the improved CoRaCMG~\cite{CoRaCMG}, restricting retrieval to commits from the same repository whose timestamps are earlier than that of the commit corresponding to the query code diff.
Future work will examine these comparisons across more generators and prompt templates.}

Human evaluation may be affected by the number, professional backgrounds, and individual judgments of the evaluators. RQ3 involves three human evaluators with software development experience: two doctoral students in software engineering and one industry expert in software development. Their interpretations of change completeness, rationale, and impact may differ. To mitigate this threat, we provide a unified scoring guide defining the five evaluation dimensions and scoring criteria of a commit message (see Section~\ref{sec:human-eval}). 
Before the formal evaluation, evaluators review representative commit messages that illustrate accurate and complete code diff descriptions, descriptions limited to local edits, and unsupported claims about change impact. They then independently score the same pilot samples and discuss disagreements about whether the messages adequately describe the code diff, explain its rationale, and accurately convey its impact. These discussions help align their interpretation of the scoring criteria. 
During formal evaluation, the three evaluators initially score commit messages independently in random order, with the source of each message concealed. We revisit substantial disagreements through discussion and retain three sets of individual ratings afterward. All reported means, pairwise Spearman correlations, and bootstrap intervals use these final ratings; the correlations therefore describe agreement after discussion. Future evaluations will include more participants from industry and open-source communities to broaden the diversity of professional backgrounds and project experience represented.

\subsection{External Validity}
External validity concerns the extent to which the findings generalize to other software projects, programming languages, LLMs, and development environments.

We evaluate CAMGen on Java, C++, and Python using two open-source LLMs and one closed-source LLM. These settings allow us to examine its effectiveness across multiple programming languages and models. However, whether the observed improvements generalize to other languages and LLMs requires further evaluation. Future work will extend the evaluation to additional languages, such as JavaScript, Go, and Rust, and include more LLMs with different code understanding and generation capabilities.

LLMs accessed through vendor APIs may change over time. We invoke DeepSeek-V3.2, DeepSeek-V4-Flash, and GPT-5.4 through vendor APIs. Providers may update server-side weights, inference strategies, safety policies, or default parameters while retaining the same model name. Consequently, identical inputs may produce different outputs over time, affecting the reproducibility of the experimental results.

\subsection{Construct Validity}
Construct validity concerns whether the operational definitions and measurements used in the experiments accurately assess the ability of CAMGen to identify core entities in code diffs.

LLM-based core-entity identification may still produce errors. CAMGen uses an LLM to identify the core entities that carry the primary change intent and uses them as shared anchors for constructing $\mathrm{View}_{\mathrm{Struct}}$ and $\mathrm{View}_{\mathrm{Func}}$. In RQ2.1, DeepSeek-V4-Pro achieves 0.8301 Precision, 0.8141 Recall, and 0.8220 F1 against the manually annotated gold labels, demonstrating strong core-entity identification performance. However, some commits may still contain missed or incorrectly identified entities, which can affect subsequent context construction. To mitigate this threat, we compare two LLMs and use the better-performing model for core-entity identification. Future work will enlarge the manually annotated dataset, refine the identification prompt, and explore methods for detecting and correcting errors in core-entity identification for code diff.

\rev{The clusters used by $\mathrm{View}_{\mathrm{Func}}$ are candidate context groups and may not coincide with functional boundaries in a repository. Shared utility code can be semantically similar across different features, while entities contributing to one feature may differ in their representations. Core-entity selection and cluster refinement focus the supplied context on the target change. The current ablation study evaluates the contribution of this cluster-selected context within CAMGen. In future work, we plan to further evaluate how well these groups capture functional relationships among code entities and how the grouping strategy contributes to commit message quality.}

\subsection{Conclusion Validity}
The threats to conclusion validity concern the transparency and reproducibility of the experimental process. In this work, the methodology and experimental design of CAMGen are detailed in Sections~\ref{sec:method} and~\ref{sec:exp}, and a replication package~\cite{dataset} has been provided to support independent reproduction. We also adopt the LLM temperature at 0.0 to reduce generation randomness. However, updates to LLMs accessed through vendor APIs may affect the exact reproducibility of the experiment results, even when the same inputs and configurations are used.

\section{Conclusions}
\label{sec:7}\label{7-conclusions-and-future-work}\label{sec:conclusion}
This study addresses the limitation that existing CMG methods provide insufficient functional context for understanding the intent and impact of code changes. We propose CAMGen, a core change-aware multi-view framework for CMG. It identifies core entities in the code changes and uses them as anchors to retrieve functional and structural context from the repository, while also retrieving similar diff-message pairs as exemplars. CAMGen then provides the combined context together with the code diff to an LLM to generate commit messages. 
\rev{Experiments on Java, C++, and Python show that CAMGen outperforms SOTA CMG baselines on most automatic evaluation metrics. Human evaluation further confirms that CAMGen generates more comprehensive and concise messages with clearer explanations of change rationale and impact. These results demonstrate the value of function-aware multi-view context for generating accurate, informative, and project-consistent commit messages.}

Future work will investigate context selection for different code change types, such as bug fixes, feature additions, and refactorings, to improve the relevance of the context used for CMG. We also plan to explore incremental context updates to reduce the cost of generating messages for successive commits. Incorporating issue reports and code review discussions will provide LLMs with additional evidence about the problems and design decisions behind code changes. We will evaluate whether this evidence helps LLMs generate higher-quality commit messages.


\section*{Data Availability}
The replication package of this work has been made available at \cite{dataset}.

\begin{acks}
This work has been partially supported by the National Natural Science Foundation of China (NSFC) with Grant Nos. 62032016 and 92582203.
\end{acks}

\bibliographystyle{ACM-Reference-Format}
\bibliography{ref}

\end{document}